\documentclass[12pt,fleqn]{article}
\usepackage{verbatim,cite,amssymb,amsmath,graphics,amsthm,subfigure}
\numberwithin{equation}{section}

\theoremstyle{definition}
\newtheorem{theorem}{Theorem}[section]
\newtheorem{proposition}{Proposition}[section]

\newtheorem{corollary}{Corollary}[section]
\theoremstyle{definition}
\newtheorem{definition}{Definition}[section]

\renewcommand{\P}{{\cal P}}

\renewcommand{\H}{{\cal H}}
\newcommand{\bs}{\boldsymbol}
\newcommand{\p}{\hat{p}}
\newcommand{\<}{\langle}
\newcommand{\be}{\begin{equation}}
\renewcommand{\>}{\rangle}
\newcommand{\s}{\sigma}
\newcommand{\sm}{\mathsf{s}}

\renewcommand{\a}{\bs{\p}j_{3}[\sm j]}
\newcommand{\apr}{\bs{\p'}j'_{3}[\sm' j']}

\newcommand{\aprr}{\bs{\p}j'_{3}[\sm' j']}

\newcommand{\am}{^-\bs{\p}j_{3}[\sm j]}
\newcommand{\ap}{^+\bs{\p}j_{3}[\sm j]}
\newcommand{\ma}{\bs{\p}j_{3}[\sm j]^-}
\newcommand{\pa}{\bs{\p}j_{3}[\sm j]^+}

\newcommand{\cm}{\bs{\p}j_{3}[\sm j]^{-}}

\newcommand{\amp}{\bs{\p}j_{3}[\sm j]^{\mp}}
\newcommand{\amd}{\bs{\p}j_{3}[\sm_{R} j]^{-}}
\newcommand{\amr}{\bs{\p}j_{3}[\sm j]^{-}}

\newcommand{\Bp}{\tilde{\cal S}\cap{\cal H}_{+}^{2}}
\newcommand{\Bm}{\tilde{\cal S}\cap{\cal H}_{-}^{2}}

\newcommand{\Bmp}{\tilde{\cal S}\cap{\cal H}_{\mp}^{2}}

\newcommand{\bk}[2]{\ensuremath{\langle #1|#2 \rangle}}

\newcommand{\kt}[1]{\ensuremath{|#1\rangle}}
\newcommand{\br}[1]{\ensuremath{\langle #1|}}
\newcommand{\kb}[2]{\ensuremath{| #1\rangle\langle #2|}}

\begin{document}
%\title{Relativistic Resonances and Decay\\ 
%I. Gamow Vectors from $S$-Matrix Poles}
\title{Time Asymmetric Quantum Theory\\
II. Relativistic Resonances from $S$-Matrix Poles.}
\author{A.~Bohm\footnote{bohm@physics.utexas.edu}  \qquad
H.~Kaldass\footnote{Present address: Deutsches Elektronen-Synchrotron, DESY, Platanenallee 6, D-15738 Zeuthen, e.mail hani@ifh.de}\qquad
S.~Wickramasekara\footnote{sujeewa@physics.utexas.edu}   \\ Physics
Department\\ The University of Texas at Austin\\ Austin, Texas 78712}
\date{} \maketitle
\begin{comment}
\begin{abstract}
A state vector description for relativistic resonances   is derived
from the first order pole of the $j$-th partial $S$-matrix at the
invariant square mass value $\sm_R=(M-i\Gamma/2)^2$ in the second
sheet of the Riemann energy surface. To  associate a ket, called Gamow
vector, to the pole, we use the generalized eigenvectors of the
four-velocity operators in place of the customary momentum eigenkets
of Wigner's Poincar\'e group representations, and we replace the 
conventional Hilbert space assumptions (or asymptotic completeness)
for the in- and out-scattering states with the new hypothesis that in-
and out-states are described by two different dense Hardy subspaces with
complementary analyticity properties.  The Gamow vectors have the
following properties:\\

-They are simultaneous generalized eigenvectors of the four velocity
operators with real eigenvalues and of the self-adjoint  invariant
mass   operator $M=(P_\mu P^\mu)^{1/2}$ with complex eigenvalue
$\sqrt{\sm_R}$.\\ - They have   a Breit-Wigner distribution in the
invariant square  mass variable $\sm$ and lead to an exactly
exponential law for the decay rates and probabilities.
\end{abstract}
\end{comment}
\begin{abstract}
Relativistic resonances and decaying states are described by representations
of Poincar\'e transformations, similar to Wigner's definition of stable
particles. To associate decaying state vectors to resonance poles
of the $S$-matrix, the conventional Hilbert space assumption
(or asymptotic completeness) is replaced by a new hypothesis
that associates different dense Hardy subspaces to the in-
and out-scattering states. Then one can separate the scattering 
amplitude into a background amplitude and one or several
``relativistic Breit-Wigner'' amplitudes, which represent the
resonances per se. These Breit-Wigner amplitudes have a precisely
defined lineshape and are associated to exponentially decaying
Gamow vectors which furnish the irreducible representation spaces
of causal Poincar\'e transformations into the forward light cone.
\end{abstract}
\maketitle


\section{Introduction}\label{sec1}

Stable quantum mechanical states are characterized by one real number -- the
energy $E_{n,j}$ or, in the relativistic case, the mass $m_{n,j}$ -- in addition
to definite values of discrete quantum numbers $n$ such as charge, isospin 
(particle species labels) and by  angular momentum or spin (parity) $j^{\pi}$. Quasistable
states $D$ are characterized by a pair of real numbers, in addition to the same discrete quantum numbers.
For these two numbers one takes either energy and width $(E_R,\Gamma)$ or energy and inverse
lifetime $(E_R,\frac{1}{\tau})$ depending upon the way these quantities can be measured. In the 
relativistic case one takes mass and width $(M_R,\Gamma)$ or mass and (inverse) lifetime 
$(M_R,\frac{1}{\tau})$. 

Lifetime $\tau$ and its inverse $R\equiv\frac{1}{\tau}$, the initial
decay rate of the decay $D\rightarrow\eta$ for any decay channel $\eta$ 
are measured by fits of the experimental counting rate $\frac{1}{N}\frac{\Delta N_\eta(t)}{\Delta t}$
 to the exponential law for the partial decay rate $R_\eta(t)$:
\begin{equation}
\label{exponential}
\frac{1}{N}\frac{\Delta N_{\eta}(t_i)}{\Delta t_i} \approx R_{\eta}(t) = R_\eta e^{-Rt}\,.
\end{equation}
(Here $R_\eta(t)$ are the theoretical partial decay rates of the decay $D\rightarrow \eta$
for any decay channel $\eta$ and $\Delta N_\eta(t_i)$ is the number of decay products
$\eta$ registered in the detector during the time interval $\Delta t_i$ around
$t_i$).

In contrast, the width $\Gamma$ is measured by fits of the cross section 
for the resonance scattering process $\eta_0\rightarrow D \rightarrow \eta$
to the Lorentzian (Breit-Wigner)
energy distribution
\begin{eqnarray}
\label{bw}
\!\!\!\!\!\!\!\!\!\!\!\!\sigma_j^{BW}(E)\sim & \left| a_j^{BW}(E)\right|^2 =
\left|\frac{r_\eta}{E-(E_R-i\frac{\Gamma}{2})}\right|^2 \sim
\frac{1}{(E-E_R)^2+(\frac{\Gamma}{2})^2}\,,\, r_\eta = \sqrt{R_{\eta 0}R_\eta}\\
\nonumber
&\text{with } 0\leq E < \infty\,.
\end{eqnarray}
\begin{comment}
\begin{equation}
\label{bw}
\sigma_j^{BW}(E)\sim \left| a_j^{BW}(E)\right|^2 = 
\left|\frac{r_\eta}{E-(E_R-i\frac{\Gamma}{2})}\right|^2 \sim
\frac{1}{(E-E_R)^2+(\frac{\Gamma}{2})^2}\,,
\end{equation}
\begin{equation}
\nonumber
\nonumber \text{with }  r_\eta = \sqrt{R_{\eta_{0}}R_\eta}\,,\, 0\leq E < \infty\,.
\end{equation}
\end{comment}
(plus usually some background term $B(E)$). Decay rate $R=\sum_{\eta}R_\eta$ and resonance
width $\Gamma$ are thus different quantities, $R=\frac{1}{\tau}$ is connected with the exponential
time evolution \eqref{exponential}, and $\Gamma$ is connected with the Lorentzian energy 
distribution. But usually one does not even distinguish between $R$ and the width $\Gamma$; one
identifies $\frac{\hbar}{\tau}\equiv R$ with $\Gamma$
\begin{equation}
\label{gamma}
\Gamma = \frac{\hbar}{\tau} \equiv R
\end{equation}
and uses the words rate and width for either of them. Often one calls a quasistable state a
resonance, if the energy $E_R$ (or resonance mass $M_R$) can be measured from \eqref{bw}.
This is the case when $\Gamma/M_R$ are $\sim 10^{-1}\cdots 10^{-4}$. And one calls the quasistable
state a decaying particle when the lifetime $\tau = 1/R$ can be determined 
experimentally from \eqref{exponential},
which is usually for values of $R/M_R\,\text{\raisebox{-1mm}{$\stackrel{<}{\sim}$}}\,10^{-8}$.

In non-relativistic physics the relation \eqref{gamma} was justified by the Wigner-Weisskopf approximation
\cite{wigner_weisskopf}, which is really not an approximation of standard quantum mechanics (based on
the Hilbert space axiom), because in Hilbert space $\H$ there is no state vector $\psi^D(t)\in\H$ whose
rate for the Born probability $P_\eta(t)$
\begin{equation}
\label{born}
R_\eta(t)=\frac{d}{dt}P_\eta(t)\equiv \frac{d}{dt}(\mbox{Tr}(\Lambda_\eta|
\psi^D(t)\>\<\psi^D(t)|))
\end{equation}
obeys the exponential law \eqref{exponential}.

Recently a non-relativistic theory which fulfills
\begin{equation}
\label{tau}
\frac{\hbar}{\Gamma} = \tau\,.
\end{equation}
has been constructed using a slight modification of the standard Hilbert space
axiom \cite{aB97}. However in non-relativistic physics there was never much
doubt that resonances are not qualitatively different from decaying states
but only quantitatively by their value of width $\Gamma = \text{rate }R$. The equality
\eqref{gamma} has recently also been confirmed in atomic physics \cite{2aa,2ab}.

In relativistic physics, the situation is quite different. Based on the perturbation theoretical
definition by the self-energy of the propagators, resonances and decaying states are considered
as complicated objects that cannot be described as an exponentially decaying state or as a state
characterized by two numbers like $(M,\Gamma)$. This has recently led to some difficulties with
the definition of mass and width of the $Z$-boson \cite{tR98}.

We shall discuss here the possibility that also {\it relativistic} resonances and decaying states 
are qualitatively the same, described by representations of Poincar\'e transformations, similarly
to Wigner's theory of relativistic stable particles \cite{wigner}. This will lead to a relativistic
Gamow vector which has a ``relativistic Breit-Wigner'' energy distribution and an exponential
decay law fulfilling \eqref{tau}. This relativistic Gamow vector and/or
the corresponding relativistic Breit-Wigner will represent the resonance per se (without
background). It provides the means to precisely define a relativistic resonance
and separate the scattering amplitude into a background and the resonance
amplitude in an unambiguous way, and therewith define the mass and width of a 
relativistic resonance.

In Section~\ref{sec2} we review the non-relativistic theory and introduce the Hardy
space hypothesis which replaces the Hilbert space axiom of quantum mechanics. Section~\ref{sec3}
reviews scattering theory and presents its minor variations based on the new 
Hardy space hypothesis. The 
relativistic theory of scattering is presented in Section~\ref{sec4}, and 
Section~\ref{sec5} gives a detailed derivation of the relativistic Gamow vector from the $S$-matrix
pole. Some of its properties are deduced here. The transformation of the relativistic
Gamow vectors under (causal) Poincar\'e transformations and its consequences are the
subject of the subsequent paper \cite{rgv2}.

\section{Alternative Phenomenological Descriptions \\ and their
	Mathematical Idealizations}\label{sec2}

	In the non-relativistic phenomenological description of
	quasistable and stable states, one has two alternative
	descriptions: the S-matrix description and the Hamiltonian
	description \cite{3a}.
 	
	In the S-matrix description the partial S-matrix element with
	angular momentum  $j$ (the $j$-th partial S-matrix $S_j(E)$)
	is an analytic function on a Riemann energy surface cut along
	the real positive axis from $E_0(=0)$ to $\infty$. The
	S-matrix is written in terms of the elastic scattering
	amplitude $a_j^{\eta_0}(E)$ and the reaction amplitude
	$a_j^\eta(E)$ as 
\begin{eqnarray} S_j^{\eta_0}(E) = 2 i
	a_j^{\eta_0}(E) + 1 \mbox{ (elastic)} \\  S_j^\eta(E) = 2 i
	a_j^\eta(E) \mbox{ (reaction)} 
\end{eqnarray} where $\eta_0$
	denotes the quantum numbers of the initial state and the
	elastic channel and $\eta$ denote the reaction channels.

	If one has a Hamiltonian given by $H=H_0 +V$, where $H$
	is selfadjoint and semi-bounded (stability of matter) by $E_0$,
	then the S-matrix and Hamiltonian description are related and the 
	scattering amplitude is given by
\begin{equation}
\label{amplitude} 
a_j^{\eta}(E) = -\pi \<E, j \ldots \eta_0 |V| E, j
	\ldots \eta^-\> 
\end{equation} where $\kt{ E, j \ldots \eta^-}$
	are eigenkets of the exact Hamiltonian with real (continuous)
	eigenvalues $E$, angular momentum $j$, and other quantum
	numbers indicated by $\ldots$. The superscript ($-$) indicates
	that they are solutions of the Lippmann-Schwinger  equation
	fulfilling outgoing boundary conditions ($-i \epsilon$ in the
	denominator). The $\kt{E, j \ldots \eta_0}$ are eigenkets of
	the free Hamiltonian $H_0$.

	A stable state is, in the S-matrix description, a pole at
	$E=E_n=-|E_n|$ on the \emph{first} sheet of the Riemann
	surface. Whereas in the Hamiltonian description, a stable
	state is an eigenvector with a real discrete eigenvalue
	corresponding to the equation,  \begin{equation}
\label{ee}		
		H\kt{E_n, j, \ldots} = E_n\kt{E_n, j, \ldots}
	\end{equation}
	
	An unstable particle, in the S-matrix description, is related
	to a pole at $E=E_R-i\Gamma/2 \equiv z_R$ on the \emph{second}
	sheet of the Riemann Surface. If there is just one resonance
	with angular momentum $j$ in the $\eta$-channel then
	\begin{equation}
\label{sdescription}
		a_j^\eta(E) = \frac{r_\eta}{E-(E_R-i\frac{\Gamma}{2})}
	+B(E); \ \ r_\eta = \sqrt{R_{\eta_0} R_\eta} \end{equation}
	where the ``physical'' values of E are $0 \leq E < +\infty$.

	In the Hamiltonian description, since the selfadjoint operator
	$H$ of (\ref{ee}) cannot have a complex eigenvalue $z_R$, one
	devises an  ``effective'', complex Hamiltonian matrix
	$H_{eff}$ and the decaying state is an eigenvector of
	$H_{eff}$ with a complex eigenvalue \begin{equation}
\label{hdescription}
		H_{eff} \kt{f} = (E_R -i \Gamma /2) \kt{f}
	\end{equation} For example, in the neutral Kaon theory, one
	has two such eigenvectors $f^{K_S}$ and $f^{K_L}$ with
	\begin{equation}
\label{Heff}
		H_{eff} f^{K_{S, L}} = (M_{S, L} - i \Gamma_{S,L} /2)
		f^{K_{S, L}}	 \end{equation} Since the definition
		(\ref{hdescription}) is mathematically problematic
		because it requires the justification of a finite
		dimensional complex submatrix for a selfadjoint
		operator $H$, the definition (\ref{sdescription}) of a
		resonance as a pole of the S-matrix has  become the
		most universal definition of a resonance state.

	A complex eigenvalue of a self-adjoint Hamiltonian, $H$ is not
	possible for a vector in the Hilbert Space $\mathcal{H}$, thus
	to obtain something like  $\kt{f}$ in (\ref{hdescription}),
	one must go outside the Hilbert Space. This should not be
	surprising  because Dirac kets (generalized eigenvectors with
	eigenvalues from the continuous real energy spectrum) are also
	not elements of $\mathcal{H}$, and to give them a
	mathematically rigorous foundation, one has to use a Rigged
	Hilbert Space(RHS) (see Appendix~\ref{r}). The ordinary Dirac
	eigenkets $\kt{E, j, \ldots, \eta}$ of $H_0$ are  usually
	taken from the RHS, $\Phi \subset \mathcal{H} \subset
	\Phi^\times$, where the realization of $\Phi$ is given by the
	Schwartz Space $S$, which is the space of smooth functions
	rapidly decreasing at infinity. This means that every vector
	$\psi_{j, \eta} \in \Phi$ (with fixed value of $j$ and the
	other quantum numbers $\eta$), can be represented according to
	the Dirac basis vector expansion (nuclear spectral theorem) in
	terms of the $\kt{E, j, \eta} \in \Phi^\times$ as
	\begin{equation}
\label{nuclear_spectral}
		\psi_{j, \eta} = \int_0^\infty dE \kt{E, j, \eta}
	\bk{E, j, \eta}{\psi_{j, \eta}}  \end{equation} where the
	energy wave functions $\bk{E, j, \eta}{\psi_{j, \eta}} \equiv
	\bk{E}{\psi} = \psi(E)$ are Schwartz space functions:
	\begin{equation} \label{Schwartz-energy} \psi \in \Phi
	\Leftrightarrow \bk{E}{\psi} \in S|_{\mathbb{R}_+}
	\end{equation}

	To obtain generalized eigenvectors that fulfill something like
	(\ref{hdescription}) one follows the same RHS
	method but one uses in place of the Schwartz Space $\Phi$,
	other spaces $\Phi_+$ and $\Phi_-$ which are realized by Hardy
	functions as explained next.

	The eigenkets of the exact Hamiltonian $H=H_0+V$ that one uses
	in scattering theory are not ordinary Dirac kets,
	i.e. elements of the dual of the Schwartz space $\Phi^\times$,
	but they are kets which also have meaning for complex values
	$E \pm i \epsilon$ with infinitesimal $\epsilon >0$. In
	scattering theory, one uses two solutions of the eigenvalue
	equation \begin{equation} \label{Lippmann-Schwinger} H \kt{E,
	j, \eta ^{\mp}} = E \kt{E, j, \eta ^{\mp}}\,,\quad 0 \leq E <
	\infty \end{equation} where the superscript $\mp$ refers to
	the $\mp i \epsilon$ in the denominator of the
	Lippmann-Schwinger equation. This indicates that $\kt{E, j,
	\eta ^\mp}$ must be continued from the real energies into an
	$\epsilon$ -strip of the complex energy plane: into the lower
	half plane for ($-$) and into the upper half plane for
	($+$). This means the complex conjugate of the wave functions,
	$\overline{\bk{^\mp E}{\psi ^\mp}} = {\bk{\psi ^\mp}{E
	^\mp}}$, must not only be a smooth function of $E$ like  in
	(\ref{Schwartz-energy}) but they must also be functions that
	have an analytic continuation into the complex energy plane,
	in particular $\bk{^- \psi}{E ^-}$ must have an analytic
	continuation into the lower half plane. Hardy functions,
	elements of   ${\H}_{\mp}^2 \cap S|_{\mathbb{R}_+}$ have
	this property. Thus we take for the energy wavefunctions of a
	scattering system 
	\begin{subequations}
	  \label{hardy_wave}
	  \begin{equation}
	    \label{hardy-wave-pm} 
	    \tag{\ref{hardy_wave}$_\pm$}
		{\bk{^\mp \psi}{E ^\mp}} \in \left. \left( {\cal{H}}_{\mp}^2 \cap
		S \right) \right|_{\mathbb{R}_+} \mbox{ which implies } \bk{^\mp E}{\psi
		  ^\mp} \in \left. \left( {\cal{H}}_{\pm}^2 \cap S \right)\right|_{\mathbb{R}_+}
	  \end{equation}
	\end{subequations}
This is a new hypothesis which replaces the Hilbert space axiom
$\<^\mp E|\psi^\mp\> \in L^2(\mathbb{R}_+)$ or the Schwartz space
assumption \eqref{Schwartz-energy}.

	The generalized eigenvectors (\ref{Lippmann-Schwinger})
	representing out ($-$) and in ($+$) plane wave solutions of
	the Lippmann-Schwinger equation are therefore kets in a pair
	of Rigged Hilbert Spaces: \begin{equation} \Phi_\pm \subset
	\mathcal{H} \subset \Phi_\pm^\times; \ \ \ \ \ \ \kt{E, j,
	\eta ^\mp} \in \Phi_\pm^\times \end{equation} This means the
	vectors $\psi^- \in \Phi_+$ are given by the Dirac basis vector
	expansion
\begin{subequations}
\label{vector_expansion} 
\begin{equation}
  \label{vector_expansion_plus}
  \tag{\ref{vector_expansion}$_+$}
  \psi^- = \sum_{j, \eta} \int_0^\infty dE \kt{E, j,
    \eta^-} \bk{^- E, j,\eta}{\psi ^-} 
\end{equation} 
	where the
	energy wave functions $\bk{^- E, j, \eta}{\psi^-}=\<^-E|\psi^-\>$ 
	are Hardy functions analytic in the upper-half plane and the vectors
	$\phi^+\in\Phi_-$ are given by the Dirac basis vector expansion
\begin{equation}
\label{vector_expansion_minus}
\tag{\ref{vector_expansion}$_-$}
\phi^+ = \sum_{j,\eta}\int dE |E,j,\eta^+\>\<^+E,j,\eta|\phi^+\>
\end{equation}
\end{subequations}
where the $\<^+E,j,\eta|\phi^+\>=\<^+E|\phi^+\>$ are elements of
$\left.\H_{-}^2\cap{\cal S}\right|_{\mathbb R_{+}}$.
	In the scattering experiment, the $\psi^-
	\in \Phi_+$ represent the out-states registered by a detector
	and  the $\phi^+ \in \Phi_-$ represent the in-states prepared
	by a preparation apparatus (e.g. accelerator), as will be discussed
	in Section~\ref{sec3}.

	In the heuristic formulation using the Lippmann-Schwinger
	equations \cite{lippmann,newton}, the precise mathematical meaning of  the out and in
	plane wave solutions is usually not stated. It is
	understood that they are to provide a means to distinguish
	between in-states $\phi^+$ prepared in the past, and
	out-states $\psi^-$ registered by the detector in the future
	after they have passed the interaction region. Such a
	distinction is meaningless in the Hilbert space where only
	time symmetric solutions of the Schrodinger or Heisenberg
	equation- given by the unitary time evolution group- are
	allowed. Thus there is a contradiction between the Hilbert
	space axiom of quantum theory and the distinction between in-
	and out- states in scattering theory. Since the solutions to
	the Lippmann-Schwinger equation  are kets $\kt{E, j, \eta
	^\mp}$, they are not elements of the Hilbert space, and one
	can choose them to be two solutions of the same eigenvalue
	equation with two different, time asymmetric, boundary
	conditions. This is precisely what we intend to do
	\cite{2b}. We replace the axiom of orthodox von Neumann
	quantum mechanics, which asserts that  the set of prepared
	in-states and the set of detected observables (or out-states)
	are both equivalent to the Hilbert space \begin{equation}
	  \label{hilbert}
	\mbox{\hspace*{-1cm}\{set of prepared in-states $\phi$\} =  \{set of
	detected out-observables $\psi$\} = $\mathcal{H}$} \end{equation} by
	a new axiom.

  	This axiom states that the prepared states, defined by the
	preparation apparatus (accelerator), are described by
	\begin{equation}
\label{phi+}
		\{\phi^+\} = \Phi_- \subset \mathcal{H} \subset
	\Phi_-^\times \end{equation} and the registered observables,
	defined by the registration apparatus (detector) are described
	by  \begin{equation}
\label{phi-}
		\{\psi^-\} = \Phi_+ \subset \mathcal{H} \subset
	\Phi_+^\times, \end{equation} where $\mathcal{H}$ in
	(\ref{phi+}) and (\ref{phi-}) denotes the same Hilbert space
	but $\Phi_-$ and $\Phi_+$ are the two different Hardy spaces which are dense
	in $\mathcal{H}$.  
%(See Appendix 1
	%(\ref{Hardy_above_realization}),
	%(\ref{Hardy_below_realization})) .  
For the non-relativistic
	case this axiom is a formulation of time asymmetric boundary
	conditions for the solutions of the time symmetric Schrodinger
	and the Heisenberg differential equations, respectively. It is
	correct, as stated in section 3.2 of \cite{weinberg}, that in-states
	$\phi^+$ and out-states $\psi^-$ do not inhabit two different
	Hilbert spaces. However, in contrast to what is implied in
	\cite{weinberg}, the new hypothesis (\ref{phi+}), (\ref{phi-})
	postulates that the in- and out- kets
	(\ref{Lippmann-Schwinger}), which are generalized eigenvectors
	and not in $\mathcal{H}$, are from {\em two} different spaces
	$\Phi_{\pm}^\times$, because (\ref{phi+}), (\ref{phi-})
	postulates that the set of in-states $\{\phi^+\}$, and the set
	of out-states $\{\psi^-\}$, are \emph{different} (dense)
	subspaces of the same Hilbert space $\mathcal{H}$. These two
	dense subspaces are the Hardy spaces $\Phi_-$ and $\Phi_+$,
	whose wave functions have different but complementary
	analyticity properties\footnote{The possibility of two distinct spaces for the prepared states
	$\{\phi^+\}$ and registered observables $\{\psi^-\}$ is
	already contained in the  historical paper of Feynman
	\cite{3c}. He distinguishes between the state at times $t' <
	t_0$ which is defined by the preparation (our prepared states
	${\phi^+}$) and what he calls the ``state characteristic of
	the experiment'' at time $t'' >t_0$ (our registered
	observables ${\psi^-}$).  He mentions the possibility  that
	$\{\psi^-\} \neq \{\phi^+\}$ in Footnote 14, attributing  it
	to H. Snyder, but does not consider it. We implement this
	possibility by the choice of the two Hardy spaces $\Phi_-$ and
	$\Phi_+$. From the hypothesis (\ref{phi+}), (\ref{phi-})  one
	derives (by Fourier transform, using the   Paley-Wiener
	theorem for Hardy Class functions) the quantum mechanical time
	asymmetry \cite{antoniou}.}.

	The axiom \eqref{phi+}, \eqref{phi-} is the only new hypothesis needed
	for a time asymmetric quantum
	theory. The new hypothesis is needed in order to obtain a
	consistent theory of resonance scattering and decay, that gives
	the Weisskopf-Wigner approximation if superpositions over the
	energy continuum are neglected \cite{aB97}.
	With this new axiom \eqref{phi+} and \eqref{phi-}, quantum
	mechanics is no longer a strictly reversible theory, it
	encapsulates time asymmetry. Except for the
	new axiom \eqref{phi+} and \eqref{phi-}, all the other basic
	assumptions (or axioms) of quantum mechanics, including the
	dynamical differential equations (the Schrodinger equation for
	the states or the Heisenberg equation for the observables)
	remain the same, but they will be extended to the  new vectors
	(kets) of $\Phi_+^\times$ and $\Phi_-^\times$. In particular,
	the Born probability for measuring the observable $\psi^-$ in
	the state $\phi^+$ is \begin{equation}
\label{bornprob}
		P_{\psi^-} (\phi^+(t)) = |(\psi^-, \phi^+(t))|^2 =
	|(\psi^-(t),\phi^+)|^2 	 \end{equation} and this axiom will be
	extended to elements of $\Phi_+^\times$.

	With the new mathematical concepts of Rigged Hilbert Spaces of
	Hardy type (\ref{phi+}) and (\ref{phi-}), we can also give a
	precise meaning to such vectors as in
	(\ref{hdescription}):

	We replace the
	phenomenological Breit-Wigner in (\ref{sdescription}) where $0
	\leq E < \infty$ by an ``exact'' Breit-Wigner for which the
	energy extends from $-\infty_{II} < E < \infty$. Then one can
	associate to it an ideal Gamow vector $\psi_j^G$, defined as
	the continuous  superposition of the Lippmann-Schwinger-Dirac
	kets $\kt{E, j \ldots ^-}$ with an ``exact'' Breit-Wigner as a
	wave function \footnote{The normalization factor $\sqrt{2 \pi
	\Gamma}$ is an inconsequential convention.}  \begin{eqnarray}
		\nonumber
		&a_j^{BW}(E) = \frac{r_\eta}{E-(E_R-i\frac{\Gamma}{2})}
		\Longleftrightarrow\\
	        \label{bwgk}
		&\psi_j^G = \sqrt{2 \pi \Gamma}
		\kt{z_R, j, \ldots ^-} = \frac{i \sqrt{2 \pi
		\Gamma}}{2\pi} \int_{-\infty_{II}}^{+\infty} dE
		\frac{\kt{E, j, \ldots^-}}{E-z_R} \\ &z_R = E_R -
		i\Gamma/2\nonumber %\ \ \ \ \ \ \ \ \ \ \ \ \ \ \ \ \ \ \ \
		%\	\ \ \ \ \ \ \ \ \ \ \ \ \ \ \ \ \ \ \ \ \ \ \
		%\ \ \ \ 
	\end{eqnarray}	 The subscript
		$II$ in $-\infty_{II}$ indicates that the analytic
		continuation has been done in the second sheet of the
		analytic S-matrix where the positions
		$z_R=E_R-i\Gamma_R/2$ of the resonance poles are
		located. The  Gamow vector $\psi_j^G$ is thus
		defined as the continuous linear superposition of the
		$\kt{E, j, \ldots^-}$ with Breit-Wigner energy wave
		functions $\bk{^- E, j}{\psi_j^G} \sim
		a_j^{BW}(E)$. In contrast to the superposition
		(\ref{vector_expansion}) for ordinary vectors
		$\psi^-$, the integration in (\ref{bwgk}) extends over
		$-\infty_{II}  < E < \infty$ and this is only possible
		if $\psi^G$ is a functional $\psi^G(\psi^-) =
		\bk{\psi^-}{\psi^G}$ over the Hardy space, $\psi^- \in
		\Phi_+$. This definition of $\psi^G$ by $(\ref{bwgk})$
		has been suggested by the continuous deformation of
		the integral for the S-matrix \cite{aB97}.

 	For the vector, $\psi_j^G \in \Phi_+^\times$ defined in
	(\ref{bwgk}) (and only if the integral extends to
	$-\infty_{II}$) one can derive (using the property of Hardy
	functions) \cite{aB97} that  \begin{equation}
\label{matrix}
		\bk{H\psi_\eta^-}{\psi^G} \equiv \br{\psi_\eta^-}
		H^\times \kt{\psi^G}   =
		(E_R-i\Gamma/2)\bk{\psi_\eta^-}{\psi^G}   \ \ \
		\forall  \  \ \psi_\eta^- \in \Phi_+ \end{equation}
		when $H=H_0+V$ is self-adjoint (and semibounded). This
		justifies the notation $\psi_j^G = \sqrt{2 \pi \Gamma}
		\kt{E_R - i \Gamma /2, j \ldots ^-}$. In Dirac's
		notation the arbitrary $\psi_j^- \in \Phi_+$ is
		omitted and (\ref{matrix}) is written as
		\begin{equation}
\label{in_the_exponent}
		H^\times \kt{E_R-i \Gamma/2, j, \ldots ^-}  =
		(E_R-i\Gamma/2)\kt{E_R-i\Gamma/2, j, \ldots ^-}
		\end{equation} Dirac also omitted the $^\times$ of
		$H^\times$ which is uniquely  defined as the extension
		of the operator $H^\dagger = H$ to $\Phi_+^\times$  by
		the first equality in (\ref{matrix}), c.f. Appendix $A$.
		The Gamow ket $\kt{E_R - i \Gamma/2 ^-}$ is thus a generalized 
		vector like a 
		Dirac ket but with complex eigenvalue $z_R$, where $z_R$
		is given by the position of the S-matrix pole of the
		resonance in the  lower half plane. However, whereas the usual
		Dirac ket, is (most of the time) thought of as a
		functional over the Schwartz space~\footnote{Physicists usually
		  do not give a mathematical definition of the Dirac ket, but, if they do
		  \cite{streater}, they define them as functionals over the space $S$ or
		  ${\cal D}$. The Lippmann-Schwinger kets must be defined over a function
		  space of analytic functions, cf. the remarks leading to \eqref{hardy-wave-pm}.}
		  $\Phi$, the Gamow
		ket (\ref{bwgk}) is an element of $\Phi_+^\times
		\supset \Phi^\times$, i.e., a functional over the Hardy space $\Phi_+$.
		The Gamow ket \eqref{in_the_exponent} and the Lippmann-Schwinger
		kets \eqref{Lippmann-Schwinger} are vectors which are ``more general''
		than the usual Dirac kets.

	The Gamow vector with exact Breit-Wigner energy distribution
	defined by (\ref{bwgk}) represents the state associated to the
	Breit-Wigner scattering amplitude (without the background
	$B(E)$ of (\ref{sdescription})) of width $\Gamma$. For this
	state vector one derives the exponential time evolution:
	\begin{equation}
\label{asymmetry}
		\psi^G(t) \equiv e^{-iH^\times t} \psi^G = e^{-i E_R
		t} e^{-\frac{\Gamma}{2}t} \psi^G; \mbox{ for } t\geq 0
		\mbox{ only.}  \end{equation} Formally
		(\ref{asymmetry}) is just (\ref{in_the_exponent})
		applied in the exponent, but for a precise derivation,
		one needs again the mathematical properties of the
		Hardy functions \cite{aB97} and the time asymmetry $t
		\geq 0$ is a consequence of this. Written in the form
		(\ref{r6}), the time evolution
		(\ref{asymmetry}) is also written as \begin{eqnarray}
\label{asymmetry2}
	\bk{e^{iHt} \psi_\eta^-}{E_R-\Gamma/2^-} \equiv
		\bk{\psi_\eta^-}{e^{-iH^\times t}|E_R-\Gamma/2^-} =
		\nonumber \\ e^{-iE_R t} e^{-\frac{\Gamma}{2}t}
		\bk{\psi_\eta^-}{E_R-i\Gamma / 2^-} \mbox{\ \ \ \ \
		$\forall$  \   $\psi_\eta^- \in \Phi_+$  \ and for \
		$t \geq 0$.}   \end{eqnarray}
	
	Since $\bk{\psi_\eta^-}{\psi^G(t)}$ represents --in analogy to
	(\ref{bornprob})--  the probability amplitude to find the decay
	product $\eta$ (described by $\psi_\eta^-$) in the state
	$\psi^G(t)$ we have derived the exponential law for the
	probabilities of a transition from the Gamow state $\psi^G$
	into any decay product $\eta$: \begin{equation}
\label{decayprobx}
	|\bk{\psi_\eta^-}{\psi^G(t)}|^2 = e^{-\Gamma t}
	|\bk{\psi_\eta^-}{\psi_j^G(0)}|^2  \mbox{,  \ \ \ \     for t
	$\geq$ 0} \end{equation} where $\Gamma$ is the width of the
	Breit-Wigner amplitude in (\ref{bwgk}), i.e., 
	$\Gamma = -2Im z_R$ where $z_R$ is the resonance pole position. This
	exponential law shows that the lifetime is given by $\tau =1/\Gamma$.

	Another result of (\ref{asymmetry2}), (\ref{decayprobx}) is
	the time asymmetry $t\geq0$.  It has been called microphysical
	irreversibility or fundamental quantum mechanical time
	asymmetry, for further discussions of this subject we refer to
	the literature \cite{gellmann, PRA, 5, ludwig, haag}.  The result (\ref{asymmetry2})
	shows that $\psi^G$ has only a semigroup time evolution
	described by the operator $U^\times(t) = e^{-iH^\times t}$ in
	$\Phi_+^\times$  (defined by (\ref{r6})). This
	is in  contrast to the unitary group evolution described by
	$U^\dagger(t) = e^{-iH^\dagger t}, \mbox{ $-\infty < t <
	+\infty$}$ for every vector in the Hilbert space
	$\mathcal{H}$.  \begin{comment}The operator  $U^\times(t)$ is the uniquely
	defined extension to the space $\Phi_+^\times$ of the operator
	$U^\dagger(t)$, but only for $t \geq 0$ \cite{aB97}.\end{comment}
An analogous result is also obtained in the relativistic case from the
transformation properties of the relativistic Gamow vectors under Poincar\'e
semigroup transformations into the forward light cone. This is the subject
of the subsequent paper~\cite{rgv2}.

\addtocounter{footnote}{1}
\section{S-Matrix$^\text{\ref{f}}$}\label{sec3}
\footnotetext{Without replicating the details here, we closely follow chapter $3$ of 
\cite{weinberg} in order to both display the analogy (and comparability)
and expose the differences 
between our development and the standard views in relativistic
quantum theory. Our notation transcribes into that of~\cite{weinberg}
as $\{\phi^{in/out}, \psi^{out}\}\rightarrow\Phi_{g}$ and 
$\{ \phi^{+},\psi^{-}\}\rightarrow \Psi^{\pm}_{g}$.
In~\cite{weinberg}, the multi-particle basis vectors
are also denoted by $\Psi^{\pm}_{\alpha}$ where $\alpha=\{p_{1}\s_{1}
n_{1},p_{2}\s_{2}n_{2},\cdots\}$, $\s$ is the third component
of the spin, and $n$ is the species label.\label{f}}
In the previous section we used heuristic arguments about the
analyticity property of the out- and in- Lippmann-Schwinger energy
wave functions $\<^-E|\psi^-\>$ and $\<^+\phi|E^+\>$ to arrive at the
two Hardy spaces $\Phi_+$ and $\Phi_-$, Their vectors
$\psi^-\in\Phi_+$ and $\phi^+\in\Phi_-$ were defined as those vectors
that could be represented by the Dirac basis vector 
expansion in terms of the out- and in-
Lippmann-Schwinger plane wave kets \eqref{vector_expansion_minus} and 
\eqref{vector_expansion_plus} using Hardy
functions. Their interpretation was already hinted at by calling
$\phi^+$ prepared states and $\psi^-$ registered observables. 

Every experiment in quantum physics can be subdivided into a preparation part and a
registration part,
\begin{comment} The preparation apparatus prepares the states
$\phi^+$ and the registration apparatus registers (or detects) the
observables $|\psi^-\>\<^-\psi|$. And the new hypothesis \eqref{phi+}
\eqref{phi-} says that states $\{\phi^+\}$ and observables $\{\psi^-\}$ are
represented by different dense Hardy subspaces of the same Hilbert
space. We apply this hypothesis to the relativistic scattering
experiment to derive the relativistic Gamow vector from the $S$-matrix
pole.
\end{comment}
though this division is not always unique, in particular for complicated experiments
\cite{ludwig}. We shall apply this principle to the typical (relativistic) scattering
experiment as depicted in Figures~$\text{\ref{scatt_fig_a}--\ref{scatt_fig_d}}$. 
Most approaches in the foundation of
quantum mechanics ascribe fundamental importance to the notions of state and of observable
and differentiate between them \cite{ludwig}.
In the standard formalism of scattering theory however, the basic entities are
the interaction-free in-states $\phi^{in}$ which become $\phi^+$ in the interaction
region, and the interaction-free out-states $\phi^{out}$ which are the $\phi^-$
in the interaction region. The in-state $\phi^+$ is the state defined by the preparation
apparatus (accelerator). The observable $|\psi^-\>\<\psi^-|$ is defined by the
registration apparatus (detector); and according to the new hypothesis \eqref{phi+}
 \eqref{phi-} states $\{\phi^+\}$ and observables $\{\psi^-\}$ 
are represented by different dense Hardy subspaces of the same Hilbert space.
The entities of quantum theory that are confronted with the experimental data
(ratios of large numbers of detector counts) are the Born probabilities
\eqref{bornprob} of an observable $\psi^-$ in a state $\phi^+$. The out-states
$\phi^-$ of standard scattering theory do neither represent apparatus controlled
states nor observables. Therefore, if one wants to distinguish 
meticulously between states and observable in scattering experiments, standard
scattering theory has to be adapted slightly, in particular the out-state $\phi^-$ 
as fundamental notion has to be replaced by an out-observable $\psi^-$.

In a scattering experiment, the experimentalist prepares 
the asymptotic state 
$\phi^{in}$ describing the non-interacting projectile and target beams. This is 
depicted in Figure~\ref{scatt_fig_a} for the (fictitious) scattering experiment
$\pi^-p\rightarrow\Lambda K^\circ$~\footnote{In realistic experiments the states are not pure but
mixtures $W^{in}=\sum w_{\alpha}|\phi^{in}_{\alpha}\>\<\phi_{\alpha}^{in}|$
and the observables are not given by one-dimensional projection operators
$|\psi^{out}\>\<\psi^{out}|$ but by
$\Lambda^{out}=\sum \lambda(\beta)|\psi_{\beta}^{out}\>\<
\psi_{\beta}^{out}|$. The pion $\pi$ and the proton $p$ are
uncorrelated and are thus described by the direct product of density
operators $|\pi^{in}\>\<\pi^{in}|\otimes|p^{in}\>\<p^{in}|$ rather
than by a vector $\phi^{in}=|\pi p^{in}\>$, but the standard discussion
in terms of vectors suffices here.}.
It is assumed that the time translation
generator $H$ can be divided into two terms, the ``free-particle''
Hamiltonian $K$($=P_{1}^{0}+P_{2}^{0}$ at rest) 
and an interaction part $V$ (or something similar):
\begin{equation}
H=K+V\,,\label{3.1}
\end{equation}
where the split of $H$ into $K$ and $V$ will be different if
different in- and out-particles are involved.

The state vectors $\phi^{in}(t)=e^{-iKt}\phi^{in}$ 
evolve in time (in the Schrodinger picture) according to the free Hamiltonian $K$.
When the particles reach their 
interaction regions, (cf. Figure~\ref{scatt_fig_a})
the free in-state vector $\phi^{in}$ changes into an exact
state vector $\phi^{+}$ whose time evolution is governed by the exact
Hamiltonian $H=K+V$. This change is usually described by the Moeller wave
operator $\Omega^+$:
\begin{equation}
\Omega^{+}\phi^{in}(t)\equiv\phi^{+}(t)=e^{-iHt}\phi^{+}=
\Omega^{-}\phi^{out}(t)\label{3.2} 
\end{equation}
When the post-interaction particles move apart the exact state vector $\phi^+(t)$ changes into the free
out state vector $\phi^{out}$, which is described by the Moeller operator $\Omega^-$ in \eqref{3.2}. 
Here $t$ is the proper time in the center-of-mass of the projectile and
target. The vector $\phi^{out}$ thus describes a
state vector  which is determined by the preparation of 
$\phi^{in}$ and by the dynamics of the scattering process. This is expressed by:
\begin{equation}
\label{3.3}
\phi^{out}=S\phi^{in},\quad S=\Omega^{-\dagger}\Omega^{+}
\end{equation}
where $S$ is the operator that describes the dynamical transformation
of the asymptotic in state $\phi^{in}$ into an asymptotic out state
$\phi^{out}$.  
The vector $\phi^{in}$ and thus $\phi^{+}$ 
are determined by the preparation apparatus (accelerator) only and thus
$\phi^+$ and $\phi^{in}$ 
represent apparatus controlled states. The preparation of $\phi^{out}$ is depicted in
Figure~\ref{scatt_fig_b}. Since it is determined by the preparation apparatus
of $\phi^{in}$ {\em and} the interaction (dynamics) described by $V$ or $S$,
it does not represent a controlled state \cite{newton}. It is not an observable
either. 

The experimentalist also builds a detector described by the observable
$|\psi^{out}\>\<\psi^{out}|$ (cf. Figure~\ref{scatt_fig_c}). For the sake of definiteness we want to
consider the reaction
\begin{equation}
\pi^-p\rightarrow\Lambda K^\circ,\
K^\circ\rightarrow\pi^-\pi^+\,.\label{3.4}
\end{equation}
The vector $\psi^{out}$ represents the asymptotically free
out-observable (usually also called out-states) which is
$|\pi^+ \pi^- \Lambda^{out}\>$ in the case \eqref{3.4}. The observable
vectors also evolve in time (in the Heisenberg picture) according to the free Hamiltonian,
but since they are observables (solutions of the Heisenberg equation) they evolve by the
adjoint (or conjugate) operator $e^{iKt}$:
\begin{equation}
\psi^{out}(t)=e^{iKt}\psi^{out}\nonumber
\end{equation}

A scattering experiment
consists of a preparation apparatus and a registration apparatus
(detector), as depicted in Figure~\ref{scatt_fig_d}.

The detector, or generally the registration apparatus,
 registers an observable $|\psi^{out}\rangle
\langle\psi^{out}|$ outside the interaction region. This observable
vector $\psi^{out}$ comes
from a vector $\psi^-$ which, in analogy to \eqref{3.2}, is given by
\begin{equation}
\label{3.5}
\psi^{-}=\Omega^{-}\psi^{out}
\end{equation}
in the interaction region. $\psi^{out}$ is in the asymptotic region what
the observable $\psi^-$ is in the interaction region, and $\phi^{in}$ is in 
the asymptotic region the state which becomes $\phi^+$ in the interaction region.
The time evolution of $\psi^-$ and $\phi^+$ is governed by the exact Hamiltonian 
\begin{equation}
\psi^-(t)=e^{iHt}\psi^-\equiv {\cal U}(t)\psi^-\qquad 
\phi^+(t)=e^{-iHt}\phi^+\equiv {\cal U}^\dagger(t)\phi^+
\label{3.5a}
\end{equation}
The observable $\psi^{out}$ is of 
course not the same as the state $\phi^{out}$, since $\phi^{out}$, like
$\phi^{+}$ and $\phi^{in}$, is prepared  by the 
accelerator and $\psi^{out}$, and $\psi^{-}$, is defined (or controlled)
entirely by the 
detector. Thus the $\phi^+$ are entirely determined by the accelerator and
the $\psi^-$ are determined by the detector only (which is not the case
for the $\phi^-\leftarrow \phi^{out}$).

Hence the set of vectors $\{\psi^{-}\}$ could be, and in our case
is, distinct from the set of vectors $\{\phi^{+}\}$. This is the 
contents of the
hypothesis \eqref{phi+}, \eqref{phi-}.

The vectors $\phi^+(t)$  fulfill the Schrodinger
equation with Hamiltonian $H$ and the vectors $\psi^-(t)$ 
fulfill the Heisenberg equation of motion. This explains
the well known difference in the sign of the exponent in
 \eqref{3.5a}. This difference is not of great importance in  Hilbert space
quantum mechanics because with the Hilbert space boundary condition
\eqref{hilbert} the Schrodinger as well as the Heisenberg equation integrates
to the unitary group solution with
\mbox{$-\infty<t<\infty$}~\footnote{This is a consequence of the
specific mathematical properties of the Hilbert space and
proven by the Stone-von Neumann theorem \cite{stone}.}. 
However, this difference in the sign of the exponent
in \eqref{3.5a} is very important for the Hardy
space boundary conditions \eqref{phi+} and \eqref{phi-}, because in the Hardy space the
dynamical differential equations integrate to be semigroup solutions of
\eqref{3.5a} with $0\leq t<\infty$. As a consequence of this one obtains the
time asymmetry $t\geq 0$ for the Born probabilities \eqref{bornprob} and, 
in the relativistic case discussed below and in \cite{rgv2}, Einstein causality
for the quantum mechanical probabilities.

The measured quantities in quantum physics are the Born probabilities,
i.e., the probabilities to register an observable
$\Lambda^-=|\psi^-\>\<^-\psi|,\ \psi^-\in\Phi_+$ in a state
$\phi^+\in\Phi_-,\ W^+=|\phi^+\>\<^+\phi|$
\begin{equation}
\label{2.17prime}
\tag{\ref{bornprob}$'$}
{P}(\phi^+)=Tr(\Lambda^-W^+)=|(\psi^-,\phi^+)|^2
\end{equation}
The Born probability amplitude $(\psi^-,\phi^+)$ is expressed using the standard notions of
scattering theory 
\eqref{3.2}, \eqref{3.3} and the new \eqref{3.5} as the matrix element of the
$S$-operator:
\begin{equation}
(\psi^-,\phi^+)=(\Omega^-\psi^{out},\Omega^+\phi^{in})
=(\psi^{out},S\phi^{in})=(\psi^{out},\phi^{out})\label{3.5b}
\end{equation}
This is essentially the statement of standard scattering theory \cite{newton} \cite{weinberg}
except that there one speaks of out-states $\phi^-$ instead of out-observables $\psi^-$.
But Born probabilities correlate observables and states, not states and other
states, and the detector is not built to the specifications of prepared states, but
to the specification of observables.

The matrix element \eqref{3.5b} is the probability amplitude for the observable
$\psi^-$ in the state $\phi^+$. It can thus also be given in terms of the asymptotic 
quantities as the probability amplitude for the observable $\psi^{out}$ (e.g.,
$\psi^{out}=|\Lambda\pi^+\pi^-\>$ in the example \eqref{3.4}) in the
``state'' $\phi^{out}$.

The vectors $\phi_{\alpha}^{+}\in\Phi_-$ 
are the prepared in-states and the detected out-observable vectors
$\psi_{\beta}^{-}\in\Phi_+$ are often also {\em called} the out-states \cite{weinberg} and the array
of complex amplitudes $(\psi_{\beta}^{-},\phi_{\alpha}^{+})$ is 
called the $S$-matrix. The labels $\alpha$ and $\beta$ stand for a whole
collection of {\it{discrete}} quantum numbers. The $S$-matrix
is also defined when $\alpha$ and/or $\beta$ are continuous labels,
only then the $S$-matrix does not
represent probability but a probability density. 
These continuous labels appear because not all quantum numbers are
discrete, in particular the scattering energy $E$ is continuous. We
obtain these continuous labels when we insert the Dirac basis vector
expansions \eqref{vector_expansion_plus} for $\psi^-$ and 
\eqref{vector_expansion_minus} for $\phi^+$ into
\eqref{3.5b}. 
\begin{equation}
(\psi^-,\phi^+)=\sum_\alpha\sum_\beta\int\int dE_\alpha
dE_\beta\<\psi^-|E_\alpha\alpha^-\>\<^-E_\alpha\alpha|E_\beta\beta^+\>
\<^+E_\beta\beta|\phi^+\>\label{3.5c}
\end{equation}
\begin{equation}
\!\!\!\!\!\!=(\psi^{out},S\phi^{in})=\sum_\alpha\sum_\beta\int\int dE_\alpha
dE_\beta\<\psi^{out}|E_\alpha\alpha\>\<E_\alpha\alpha|S|E_\beta\beta\>
\<E_\beta\beta|\phi^{in}\>\label{3.5d}
\end{equation}
In the second expansion we have used the Dirac basis expansion with
respect to eigenkets of the free Hamiltonian $K$ of \eqref{nuclear_spectral} and the
same for the observables. For instance if the quantum numbers are
\begin{equation}
E_\alpha,\alpha=E_\alpha, j, j_3, n\label{3.5e}
\end{equation}
then the two complete systems of commuting observables are
\begin{equation}
K,\bs{J}^2,J_3,N\qquad H,\bs{J}^2,J_3,N\label{3.5f}
\end{equation}
(by $N$ we denote the operator of the particle label quantum
number $n$ e.g., charge operators)

In the mathematically heuristic formulation \cite{newton}
(i.e., where the kets $|E_\alpha\alpha^\pm\>$
and $|E_\alpha\alpha\>$ are not mathematically defined as functionals)
one obtains the identities~\footnote{From the 
Lippmann-Schwinger equation
\begin{equation}
\label{lipp_sch}
\tag{\ref{3.5ga}$a$}
|E_\alpha^\pm\>=\left(1+\frac{1}{E_\alpha-H\pm
 i0}\right)|E_\alpha\>\equiv\Omega^\mp |E_\alpha\>
\end{equation}
and from $\Omega^{-\dagger}\Omega^{-}=1$ one obtains
$\<\psi^-|E^-\>=\<\Omega^-\psi^{out}|\Omega^-|E\> =\<\psi^{out}|\Omega^{-\dagger}\Omega^-|E\>
=\<\psi^{out}|E\>$ and similar for $\Omega^{+\dagger}\Omega^{+} = 1$. But $\Omega^{-\dagger}$
and $\Omega^{-}$ must somehow be defined as conjugate operators $(\Omega^-)^\times$ and
$((\Omega^-)^\times)^-$ in some spaces; see also \cite{gadella}.}:
\begin{equation}
\<E_\alpha\alpha|S|E_\beta\beta\>=\<^-E_\alpha\alpha|E_\beta\beta^+\>\label{3.5g}
\end{equation}
and
\begin{equation}
\<E|\psi^{out}\> = \<^-E|\psi^-\>\,;\quad\<E|\phi^{in}\>=\<^+E|\phi^+\>\label{3.5ga}
\end{equation}

Since we are here not interested in the problem of expressing the
exact $|E,\alpha^\mp\>$ in terms of the interaction free $|E,\alpha\>$,
e.g., by a perturbation series for the Lippmann-Schwinger equation \eqref{lipp_sch},
we do not want to work here any further with the asymptotically free
eigenstates $|E,\alpha\>$. We shall only work with the interaction 
incorporating states $\phi^+$ and $\psi^-$ and the basis vectors
$|E,\alpha^\mp\>$ which are basis vectors of the representation spaces
for the exact Poincar\'e transformations \cite{weinberg}. In this spirit, we shall
use the equality \eqref{3.5g} only as a definition of the $S$-matrix on the
left hand side in terms of the exact eigenvectors $|E,\alpha^-\>$ and
$|E,\beta^+\>$ on the right hand side. And we shall use the equalities \eqref{3.5ga}
only to state that the energy wave functions $\<^-E|\psi^-\>$, $\<^+E|\phi^+\>$,
(which by hypothesis \eqref{phi+}, \eqref{phi-} are mathematically described by
Hardy functions analytic in the upper and lower energy plane, respectively) are physically
interpreted as the energy distribution of the observable $\psi$ and of the
state $\phi$ in the asymptotic regions.

\begin{comment}
In our interpretation based on hypothesis \eqref{phi+}, \eqref{phi-} the
prepared in-state $\phi^+$ is controlled by the preparation apparatus
for $\<E|\phi^{in}\>$ at the asymptotic in- region and the registered
out-observable $\psi^-$ is  controlled by the registration
apparatus (detector) for $\<E|\psi^{out}\>$ at the asymptotic out-region. 
\end{comment}
Thus in terms of energy wave functions, \eqref{phi+} and \eqref{phi-} say:
\begin{equation}
\label{3.13}
\<^-E|\psi^-\> \equiv \<E|\psi^{out}\> \in \left.{\cal S}\cap{\H_{+}^{2}}\right|_{\mathbb{R}_+}
\end{equation}
\begin{equation}
\label{3.14}
\<^+E|\phi^+\> \equiv \<E|\phi^{in}\> \in \left.{\cal S}\cap{\H_{-}^{2}}\right|_{\mathbb{R}_-}
\end{equation}
where
\begin{equation}
\label{3.13a}
\tag{\ref{3.13}$a$}
\left|\<E|\psi^{out}\>\right|^{2} = \text{energy resolution of the detector in the asymptotic region}
\end{equation}
and
\begin{equation}
\label{3.14a}
\tag{\ref{3.14}$a$}
\left|\<E|\phi^{in}\>\right|^{2} = \text{energy  distribution of the prepared beam in the asymptotic
region}
\end{equation}
The $\<^-E|\psi^-\>$, $\<^+E|\phi^+\>$ describe thus the particular experimental setup
and \eqref{phi+}, \eqref{phi-} is the axiom that states that the choice of allowed energy 
wave function is (much) more limited than under the standard axiom \eqref{hilbert}.

In our interpretation based on hypothesis \eqref{phi+} \eqref{phi-} the
prepared in-state $\phi^+$ is controlled by the preparation apparatus
for $\<E|\phi^{in}\>$ at the asymptotic in- region and the registered
out-observable $\psi^-$ is   controlled by the registration
apparatus (detector) for $\<E|\psi^{out}\>$ at the asymptotic out-region. 

In addition to the $\phi^{in}$ and $\psi^{out}$ we mentioned above also
the out-state $\phi^{out}=S\phi^{in}$. This out-state is uncontrolled. Therefore
the out-going wave $\<E|\phi^{out}\>$ would be uncontrolled. It would be a
state in which an observable $|\psi^{in}\>\<\psi^{in}|$ could have
been measured in the distant past (asymptotic in-region). Such
entities have no physical meaning because a state must be prepared
first before an observable can be measured in it (causality). Thus,
\eqref{3.13} \eqref{3.14} is all that is needed for (causal) physics. It is roughly
half of what one conventionally uses~\footnote{It excludes the version of \cite{newton} (page $188$)
which is called not appealing to our physical intuition with regard to the notion of
causality, and which we here exclude by axiom \eqref{phi+}, \eqref{phi-}}.

The statement \eqref{3.13}, \eqref{3.14} for the wave functions does not tell us anything
about the spaces of the $\psi^{out}$, $\phi^{in}$ and $|E\>$, and we
do not need  them. It is sufficient to think of the in-states
$\phi^+\in\Phi_-$ and the out-observables $\psi^-\in\Phi_+$, and 
the elements $|E,\alpha^\pm\>$ of the
conjugate spaces $\Phi_\mp^\times$. \eqref{3.13} \eqref{3.14} tell us that their energy wave
function can be prepared or determined in the asymptotic region, and
\eqref{3.5g} tells us that the dynamics (particle interaction), encapsulated
in the $S$-matrix, is given by the matrix elements of the two eigenkets
$|\alpha,E^-\> = \Phi^\times_{+}$ and $|\beta,E^+\>\in \Phi^\times_-$ of the
exact Hamiltonian $H$ with the same eigenvalue, \eqref{nuclear_spectral},
and different boundary conditions.

\begin{comment}
It is more intuitive to use for the $S$-matrix 
the notation $\<E_\alpha\alpha|S|E_\beta\beta\>$ but this has no other
meaning than the one given by \eqref{3.5ga}; the ?? in \eqref{?}, where
$\<\psi^-|E^-\>\<E^+|\phi^+\>$ are Hardy functions in the lower
complex energy plane in the second (or lower) sheet of the Riemann
energy surface.
\end{comment}

The objects of our theory are thus the interaction incorporating
states $\phi^+$, observables $\psi^-$ and the corresponding kets
$|E^\pm\>$ which are eigenvectors of the full Hamiltonian $H$. The
interaction is encapsulated in the $S$ operator or the $S$-matrix \eqref{3.5g}, of
which we do not have any specific knowledge unless we know (in the
non-relativistic case)  $V$ and can calculate it from \eqref{amplitude}. 

With the use of symmetry conditions  we can reduce the $S$-matrix
in \eqref{3.5c} \eqref{3.5d} 
to a (much) smaller number of reduced matrix elements and then simplify
the integrals on the r.h.s of \eqref{3.5d}. In particular,
from energy conservation and angular momentum conservation it follows
\begin{equation}
\label{reduceds}
\<E'j'j'_3n'|S|Ejj_3n\>=\delta(E'-E)\delta_{j'_3j_3}\delta_{j'j}
\<n'||S_j(E)||n\>
\end{equation}
so that \eqref{3.5d} becomes
\begin{equation}
\label{reducedsa}
%\tag{\ref{reduceds}$a$}
(\psi^-,\phi^+)=\sum_j\int
dE\sum_{j_3}\sum_{nn'}\<\psi^-|Ejj_3n'^-\>\<^+Ejj_3n|\phi^+\>
S_j^{n'n}(E)
\end{equation}
and our non-knowledge is now encapsulated in the much smaller number
of the $j$-th partial matrix elements $S_j^{n'n}(E)$ (some of the
dependence on the $n'$ and $n$ can in a similar way be reduced using
intrinsic (e.g., isospin) symmetries).

One defines a resonance with spin
$j$ by the pole of the $j$-th partial $S$-matrix element 
$S_j(E)$ on the lower complex semiplane of the 
second sheet. From this definition one obtains -- starting from the
exact basis vectors with real energy -- the state vector of the resonance,
which we called  Gamow vectors (or Gamow kets) \cite{aB97}.
In the relativistic case symmetry properties (with respect to
interaction-incorporating Poincar\'e transformations) are even more
important than in the non-relativistic case~\footnote{In non
relativistic quantum mechanics one has some model Hamiltonian for
which one can solve equation \eqref{hdescription} with a $V$ that includes the
interaction causing decay (e.g., square well \cite{rafa}). In the
relativistic case one always starts with the free $V\rightarrow0$
asymptotic solution $|Ejn\>$ and constructs ``exact'' solutions
$|Ejn^-\>$ in terms of the asymptotic $|Ejn\>$ using perturbation
theory, which does not seem to work for resonances, cf.~remark 
in \cite{passera}.}. We shall now turn to the derivation of the relativistic
Gamow vector from the first order pole of the relativistic $S$-matrix.

\section{Relativistic Scattering and the Definition of the Relativistic In- and Out-Plane Waves}
\label{sec4}

The relativistic theory of resonances does not require the
non-relativistic theory as a backdrop but can be developed from the
relativistic $S$-matrix theory in an analogous way. All that has to be done is to use in
place of the non-relativistic angular momentum vectors of \eqref{nuclear_spectral} and
\eqref{vector_expansion_plus}, relativistic basis vectors which span the
representation spaces of Poincar\'e transformations
\begin{equation}
\label{4.1}
|\,Ejj_{3}n\rangle  
\rightarrow |\bs{\p}j_{3}[\sm j]n\rangle\quad \text{where $\sqrt{\sm}=E^{cm}$ is the center of 
mass energy.} \\
\end{equation}
\begin{equation}
\label{4.2}
|Ejj_{3}n^{\pm}\rangle\rightarrow
|\bs{\p}j_{3}[\sm j]n^{\,\pm}\rangle
\end{equation}
(Here we use not the momentum $\bs{p}$ but the space component of the $4$-velocity
$\bs{\p}=\bs{p}/m$ as additional label of the relativistic basis kets, cf. below after
\eqref{4.17} and \cite{hani}.

The two sets of bases are again thought of as being related by the Moeller wave operators
\begin{equation}
\label{4.3}
|\bs{\p}j_{3}[w=\sqrt{\sm},j]^{\,\pm}\,\rangle
=\Omega^{\pm}|\,\bs{\p}j_{3}[w=\sqrt{\sm},j]\,\rangle\,.
\end{equation}

In the non-relativistic case one usually separates off the center of
mass motion and ignores it; this has been done in the preceding sections. Including
the center of mass motion in the non-relativistic case, one has 
\mbox{$|\bs{p} E jj_3n\>=|\bs{p}\>\otimes|Ejj_3n\>$} in place of $|Ejj_3n\>$, where
$\bs{p}$ is the center of mass momentum. In the relativistic case this
cannot be done,  the center of mass energy squared (or invariant
mass squared) $\sm$ is related to energy and momentum by
\begin{equation}
\sm^2=E^2-\bs{p}^2
=(p_1+p_2+\cdots+p_N)_\mu(p_1+p_2+\cdots+p_N)^\mu\label{4.4}
\end{equation}
where $p_1,p_2,\cdots$ are the 4-momenta of the $N$ particles involved
in the scattering process.

The vectors on the r.h.s.\ of \eqref{4.1} are the
basis vectors of a representation of Poincar\'e transformations which
are irreducible for fixed values of $[\sm j]$. For the sake of
definiteness, we restrict ourselves to two particle scattering (with
possible formations of a resonance $R$)
\begin{equation}
a+b\rightarrow R\rightarrow c+d
\label{4.5}
\end{equation}
The non-interacting in-states $\phi^{in}$ are characterized by their (two-) particle
contents which is described by the two particle space given by the direct product of 
one-particle spaces $\H^a$, $\H^b$. The latter are irreducible representation spaces of the
Poincar\'e group $\H^a = \H(m_a,j_a,n_a)$, where $m_a$ $j_a$ are mass and spin
and $n_a$ is the particle species label of particle $a$.
\begin{equation}
\H^{in}=\H^a\otimes\H^b=\H(m_aj_an_a)\otimes\H(m_bj_bn_b)
\label{4.6}
\end{equation}
The basis vectors are product basis vectors
\begin{equation}
|a\>\otimes|b\>=|\bs{p_a}j^a_3[m_aj_a]n_a\>
\otimes|\bs{p_b}j^b_3[m_bj_b]n_b\>\label{4.7}
\end{equation}
The same holds for the detected out states $\psi^{out}$
\begin{equation}
\H^{out}=\H^c\otimes\H^d=\H(m_cj_cn_c)\otimes\H(m_dj_dn_d)\label{4.8}
\end{equation}
and their basis vectors
\begin{equation}
|c\>\otimes|d\>=|\bs{p_c}j^c_3[m_cj_c]n_c\>
\otimes|\bs{p_d}j^d_3[m_dj_d]n_d\>\label{4.9}
\end{equation}
In place of labeling the basis of $\H^{in}$ by the momenta
$\bs{p}_a$, $\bs{p}_b$ and spins $j_a$, $j_b$, etc., of the
individual particles, one can combine the two vectors $|a\>$ and
$|b\>$ into an eigenvector of the total $4$-momentum operator
\begin{equation}
P_\mu^{free}=P_\mu^a+P_\mu^b\label{4.10}
\end{equation}
and the total angular momentum $j$.

This means one takes the direct product of the irreducible single
particle representations $[m_aj_a]$ and $[m_bj_b]$ of the Poincar\'e group 
and reduces it in
terms of irreducible representations of the Poincar\'e transformations
labeled by $[\sm j]$~\footnote{This is done in analogy to the
irreducible representations of the rotation group when one couples two
angular momenta $j_a$ and $j_b$ to the total angular momentum $j$ with
$|j_a-j_b|\leq j\leq|j_a+j_b|$ which is accomplished by the $SO(3)$
Clebsch-Gordon coefficients. For the Poincar\'e transformations one
needs the Clebsch-Gordon coefficients of the Poincar\'e group which
are more complicated but also known \cite{wightman}.}.

In this way one obtains for $\H^{in}$ (and similarly for $\H^{out}$
with $a,b\rightarrow c,d$) the basis vectors
\begin{equation}
|\bs{p}j_3[\sm j] l s n_an_b\>\label{4.11}
\end{equation}
where $p_\mu=(p_a+p_b)_\mu$ is the eigenvalue of $P_\mu^{free}$ in the space \eqref{4.6} and
$\sm=p_\mu p^\mu$ is the center of mass energy squared \eqref{4.4}. $j$
is the total angular momentum and $j_3$ is its third component. For a
given  $[m_aj_a]$ and $[m_bj_b]$, the  $[\sm j]$ of the vectors \eqref{4.11} that
span $\H^{in}$ take the following values:
\begin{subequations}
\label{val}
\begin{equation}
(m_a+m_b)^2\leq\sm<\infty\,;\label{val_a}
\end{equation}
and
\begin{equation}
\label{val_b}
\begin{split}
& j=0,1,2,\cdots,|j_a+j_b|\ \text{if
$j_a+j_b=$ integer}\\
&j=1/2,3/2,\cdots,|j_a+j_b|\ \text{if
$j_a+j_b=$ half-odd-integer}
\end{split}
\end{equation}
\end{subequations}
The same value $[\sm j]$ can occur in the combination of particle
$a=[m_aj_a]$ and particle $b=[m_bj_b]$ (for fixed values of $n_a$ and
$n_b$) more than once. This multiplicity is labeled by the degeneracy
quantum numbers $l$ and $s$ which take the values
\begin{equation}
s=j_a+j_b, j_a+j_b-1,\cdots,|j_a-j_b|\label{4.13}
\end{equation}
\begin{equation}
l=j+s,j+s-1,\cdots,|j-s|\label{4.14}
\end{equation}
These degeneracy quantum numbers $l$, $s$, which distinguish the identical irreducible
Poincar\'e group representations $[\sm j]$, have the interpretation of total
orbital angular momentum $l$ and total spin $s$, respectively.

We shall take these two-particle vectors \eqref{4.11} with a specific Poincar\'e
transformation property $[\sm j]$, as the asymptotically
free in-state basis vectors \eqref{4.1}, with the two particle
label $n=n_a,n_b$. The general asymptotic in-state is the continuous
superposition of \eqref{4.11}
\begin{equation}
\phi^{in}=\sum_j\int_{(m_a+m_b)^2}^\infty
d\sm\sum_{l,s}\int\frac{d^3p}{2p^0}|\bs{p}j_3[\sm
j]lsn_an_b\>\<\bs{p}j_3[\sm j]lsn_an_b|\phi^{in}\>\label{4.15}
\end{equation}
This reduction of the direct product representations of the Poincar\'e group
into its irreducible components $[\sm j]$ is discussed in details in \cite{wightman}.
A similar result holds for the asymptotic out-state $\psi^{out}$. If the
incoming (outgoing) particles have $j_a=j_b=0$ ($j_c=j_d=0$), then
$s=0$ and $l=j$, and there is no degeneracy. We shall therefor in the following
suppress the degeneracy labels $l,s$ (or consider the case
$j_a=j_b=0$) and use the notation of \eqref{4.1}.

Instead of using as basis vectors of the irreducible representation
spaces of $[\sm j]$ the momentum eigenvectors
\begin{equation}
|\bs{p}j_3[\sm j]\>\ \text{and}\ |\bs{p}_aj_{a3}[m_aj_a]\>\ \text{same
 for}\ b,c,d\label{4.16}
\end{equation}
one could as well use eigenvectors of the 4-velocity
\begin{equation}
\p_\mu=\frac{p_\mu}{\sqrt{\sm}},\ \sm=p^\mu p_\mu;\quad
\p_{a\mu}=\frac{p_{a\mu}}{m_a},\ \text{etc.,}\label{4.17}
\end{equation}
as  basis vectors of an irreducible representation $[\sm j]$. These
4-velocity basis vectors $|\bs\p j_3[\sm j]\>$ have the property
\begin{equation}
\hat{P}_\mu|\bs\p j_3[\sm j]\>=\p_\mu|\bs\p j_3[\sm j]\>\label{4.18}
\end{equation}
where 
\begin{equation}
\hat{P}_\mu=P_\mu M^{-1},\quad M=(P_\mu P^\mu)^{1/2}\label{4.19}
\end{equation}
is the 4-velocity operator and $M$ is the invariant square mass
operator whose eigenvalue is the invariant mass $\sqrt{\sm}$. The
normalization of the velocity basis kets is chosen to be
\begin{equation}
\<\bs\p' j'_3[\sm' j']|\bs\p j_3[\sm
j]\>=2\p^0\delta(\sm'-\sm)\delta^3(\bs\p'-\bs\p)\delta_{j_3j'_3}\delta_{jj'}
\label{4.20}
\end{equation}
and the integration measure for this normalization is
$\frac{d^3\p}{2\p^0}$, in place of the $\frac{d^3 p}{2p^0}$ of
\eqref{4.15}. 

The reduction of the direct product of the two irreducible
representations $[m_aj_a]$ and $[m_bj_b]$ into a direct sum of
irreducible representations $[\sm j]$:
\begin{equation}
\H([m_aj_a])\otimes\H([m_bj_b])=\sum_j\int_{(m_a+m_b)^2}^\infty
d\sm\H([\sm j])\label{4.21}
\end{equation}
and the calculation of the Clebsch-Gordan coefficients can be done 
in terms of the velocity basis vectors \eqref{4.18}
in the same way as has been done in \cite{wightman} for the momentum basis vectors,
this has been shown in \cite{hani}. 

The velocity kets $|\bs\p j_3[\sm j]\>$
provide as valid a basis for the representations of the Poincar\'e
group as Wigner's momentum kets. Moreover, the 4-velocity kets are often
more useful for physics reasons because the four-velocity 
operators \eqref{4.19} may commute
with intrinsic symmetries when the four-momentum does not. Further,
 the 4-velocities seem to
fulfill to a rather good degree ``velocity super selection
rules'' which the momenta do not \cite{velocityvectors}. Our use of
velocity kets for relativistic resonances was motivated by a remark of Zwanziger \cite{zwanziger}.
The use of velocity kets
 will become important when we analytically extend the basis
vectors from the values \eqref{val_a} (the ``physical'' values of the
scattering energies) into the complex plane, as explained below.

Although the interaction free theory can be some guidance for how the
exact theory is to be constructed, the transition from the
interaction-free theory into the exact theory incorporating
interactions is not entirely clear. But we shall use the
correspondence between the free and exact theories that is usually
assumed to be provided by a pair of Moeller operators $\Omega^\pm$:
\begin{eqnarray}
&\text{interaction free theory}&\,\,\,\text{exact theory}\nonumber\\
\text{in-states}&\phi^{in}&\,\,\,\phi^+=\Omega^+\phi^{in}\nonumber\\ 
\text{out-observables}&\,\,\,\psi^{out}&\,\,\psi^-=\Omega^-\psi^{out}\nonumber\\
\text{basis vectors}&|\bs{\p} j_3[\sm j]\rangle&\,\,\,|\bs{\p} j_3[\sm
j]^\pm\rangle=\Omega^\pm|\bs{\p} j_3[\sm j]\rangle\nonumber\\ 
\text{generators}&P_\mu^{free},\ J_{\mu\nu}^{free}~\footnote{}&\,\,\,P_\mu,\ J_{\mu\nu}\nonumber\\
\label{4.22}
\end{eqnarray}
\footnotetext{This
$P_\mu^{free}$ is the generator in \eqref{4.10} when there is no interaction
between particles $a$ and $b$.}
From now on, based on the arguments in Section~\ref{sec3}, we will only 
be concerned with the exact quantities $P_\mu$, $J_{\mu \nu}$, etc.
The exact basis vectors are chosen to be eigenvectors of the complete
system of commuting observables (csco): 
\begin{equation}
\hat{P}_\mu=P_\mu M^{-1},\
M=(P_\mu P^\mu)^{1/2},\ -\hat{w}^\mu\hat{w}_\mu,\
U(L(\hat{p}))\hat{w}_3 U^{-1}(L(\hat{p}))\,,
\label{4.23}
\end{equation}
where $\hat{w}^\mu = \epsilon^{\mu\nu\rho\sigma}\hat{P}_\nu J_{\rho \sigma}$.
The eigenvalues of the csco \eqref{4.23} are
\begin{equation}
\p_\mu,\ \sm,\ j(j+1),\ j_3\label{4.24}
\end{equation}
The Dirac basis vector expansion for the free states $\phi^{in}$ and 
$\psi^{out}$ is as in \eqref{4.15}. In the theory that incorporates
interactions, the Dirac basis vector expansion for the in-state
$\phi^+\in\Phi_-$ is

\begin{equation}
\label{4.25}
\phi^+=\sum_{jj_{3}}\int_{\sm_0}^{\infty}d\sm
\int\frac{d^{3}\p}{2\p^0}|\pa\>\<^+\bs{\p}j_3[\sm j]|\phi^+\>
\end{equation}
and for an out-observable $\psi^-\in\Phi_+$ it is:
\begin{equation}
\label{4.26+}
\psi^-=\sum_{jj_3}\int_{\sm_0}^\infty d\sm\int 
\frac{d^{3}\p}{2\p^0}|\ma\>\<^-\bs{\p}j_3[\sm j]|\psi^-\>\,,
\end{equation}
\begin{equation}
\nonumber
\text{     where }\p^0 = \hat{E}(\bs{\p}) = \sqrt{1+\bs{\p}^2}\,.
\end{equation}
In \eqref{4.25} and \eqref{4.26+}, the wave functions 
of $\phi^+$ and $\psi^-$, i.e., their components  
$\<\ap|\phi^+\>$ and $\<\am|\psi^-\>$ along the basis vectors,
are functions of the continuous variables $\sm$ and $\bs{\p}$.

To simplify the notation we often omit the labels $\p, j_3,\
\eta=\p,j_3,l,s,n$ (where $n$ are degeneracy and particle species
labels) and write the multiple dimensional projection operator on the
irreducible representation space of $[\sm j]$ as
\begin{equation}
\sum_{j_3}\int\frac{d^3\p}{2\p^0}|\bs{\p}j_3[\sm
j]^\mp\>\<^\mp\bs{\p}j_3[\sm j]|=|[\sm j]^\mp\>\<^\mp[\sm
j]|=|\sm^\mp\>\<\sm^\mp|\label{4.27}
\end{equation}
Then the basis vector expansion \eqref{4.25} can be written as
\begin{equation}
\label{4.25a} 
\tag{\ref{4.25}a}
\phi^+=\sum_j\int_{\sm_0}^\infty d\sm |[\sm j]^+\>\<^+[\sm
j]|\phi^+\>=\int_{\sm_0}^\infty d\sm|\sm^+\>\<^+\sm|\phi^+\>
\end{equation}
and \eqref{4.26+} for the out-observable $\psi^-$ is written as
\begin{equation}
\label{4.26-}
\tag{\ref{4.26+}$a$} 
\psi^-=\sum_j\int_{\sm_0}^\infty d\sm |[\sm j]^-\>\<^-[\sm
j]|\psi^-\>=\int_{\sm_0}^\infty d\sm|\sm^-\>\<^-\sm|\psi^-\>
\end{equation} 
For the wave function, now considered just as a function of the center
of mass energy squared $\sm$, we thus use the notation
\begin{equation}
\label{4.25b} 
\tag{\ref{4.25}b}
\<^-\bs{\p}j_3[\sm j]|\psi^-\>\rightarrow\<^-\sm|\psi^-\>
\end{equation}
\begin{equation}
\label{4.25c} 
\tag{\ref{4.25}c}
\<^+\bs{\p}j_3[\sm j]|\phi^+\>\rightarrow\<^+\sm|\phi^+\>\,.
\end{equation}
We also write for the basis vectors
\begin{equation}
\label{4.25d} 
\tag{\ref{4.25}d}
|\bs{\p}j_3[\sm j]^\mp\>\rightarrow|\sm^\mp\>\in\Phi^\times_\pm
\end{equation}

We shall make the hypothesis that these wave functions have the
same analyticity properties in the invariant mass squared 
$\sm=(E^{\rm{cm}})^{2}$ as the energy wave functions
in~\eqref{3.13} \eqref{3.14}. 
However, in the relativistic case, due to the 
mathematical requirement of the invariance of the subspaces $\Phi_\mp$
under the action of the generators of the Poincar\'e group, 
we cannot use exactly the Schwartz space ${\cal S}$ of \eqref{3.13} \eqref{3.14}
but have to consider a 
closed subspace $\tilde{\cal S}$ of ${\cal S}$. 
This subspace $\tilde{\cal S}$, 
constructed in~\cite{sujeewa}, 
is the space of Schwartz functions which vanish at
zero faster than any polynomial.
This requirement also assures that the zero mass states do not
contribute to the Gamow vector. This avoids
the difficulty that the four velocity operators, which is centrally
significant to our construction of Gamow vectors, cannot be 
meaningfully defined in the zero-mass case in any obvious way.
The features of the space $\tilde{\cal S}$
which are needed for the construction of the relativistic Gamow
vectors are as follows: 

\begin{proposition}\label{Property2.1}\cite{sujeewa}
The triplets
\begin{equation}
\label{4.28}
\Bmp|_{{\mathbb R}_{{\sm}_0}}
\subset 
L^{2}({\mathbb R}_{{\mathsf s}_0})
\subset
\left(\Bmp|_{{\mathbb R}_{{\sm}_0}}\right)^{\times}
\end{equation}
form a pair of Rigged Hilbert Spaces.
This means the validity of the Dirac basis vector expansion
\eqref{4.25} and \eqref{4.26+} is assured.

\end{proposition}
\noindent In~\eqref{4.28}, $\mathbb{R}_{\sm_{0}}$ is the set of
physical values of the scattering energy $\sm$ for the 
process \eqref{4.5}
$\mathbb{R}_{\sm_0}=[(m_1+m_2)^2,\infty)$.
\begin{proposition}\label{Property2.2}\cite{sujeewa}
The space ${\tilde{\cal S}}$ is endowed with a nuclear Fr\'echet topology
such that multiplication by ${\sm}^{\frac{n}{2}}$,
$$
\sm^{\frac{n}{2}}\,:\,\,\tilde{{\cal S}}\cap{\cal H}_{\pm}^{2}
\rightarrow \tilde{{\cal S}}\cap{\cal H}_{\pm}^{2}\,,\quad
n=1,2,3,\cdots
$$
is a continuous linear operator in the topology of $\tilde{\cal S}$ .
\end{proposition}
Thus the relativistic characterization of $\Phi_\pm$ 
analogous to~\eqref{3.13} and~\eqref{3.14} is:
\begin{subequations}
\label{4.29}
\begin{equation}
\label{4.29a}
\psi^-\in\Phi_+\quad\text{if and only if  }
\<\am|\psi^-\>\in\Bp|_{\mathbb{R}_{\sm_0}}
\times{\cal S}(\mathbb{R}^3)
\end{equation}
\begin{equation}
\label{4.29b}
\phi^+\in\Phi_-\quad\text{if and only if  }
\<\ap|\phi^+\>\in\Bm|_{\mathbb{R}_{\sm_0}}
\times{\cal S}(\mathbb{R}^{3})\,;
\end{equation}
\end{subequations}
where ${\mathbb R}^{3}$ is the space of components of the
$4$-velocity
and the Hilbert space $\H$ of~\eqref{phi+} \eqref{phi-} is realized by the
function space
\begin{equation}
\label{4.30}
L^{2}({\mathbb R}_{\sm_0},d\sm)\times L^{2}\left({\mathbb R}^3,
\frac{d^{3}\p}{2\p^0}\right)\,.
\end{equation}
In the truncated notation of \eqref{4.25a} and \eqref{4.26-}, the
definition \eqref{4.29a} and \eqref{4.29b} of the space $\Phi_-$ of
prepared in-states $\phi^+$ and of the space $\Phi_+$ of the detected
out-observables $\psi^-$ is
\begin{equation}
\label{4.29c}
\tag{\ref{4.29}c}
\psi^-\in\Phi_+\ \text{in and only if}\ \<^-\sm|\psi^-\>\in\tilde{\cal
S}\cap\H^2_+|_{{\mathbb{R}}_{\sm_0}}
\end{equation}
\begin{equation}
\label{4.29d}
\tag{\ref{4.29}d}
\phi^+\in\Phi_-\ \text{in and only if}\ \<^+\sm|\phi^+\>\in\tilde{\cal
S}\cap\H^2_-|_{{\mathbb{R}}_{\sm_0}}
\end{equation}  

In mathematics one calls the space of Lebesgue square integrable
functions $L^2({\mathbb{R}}_{\sm_0},d\sm)$ a realization of the
abstract Hilbert space. In the same way we will call the triplet of
function spaces
\begin{equation}
\label{4.30a}
\tag{\ref{4.30}a}
\tilde{\cal S}\cap\H_\pm^2|_{{\mathbb{R}}_{\sm_0}}\otimes{\cal S}({\mathbb{R}}^3)
\subset L^2({\mathbb{R}}_{\sm_0},d\sm)\otimes L^2({\mathbb{R}}^3,\frac{d^3\p}{2\p^0})
\subset\left(\tilde{\cal S}\cap\H_\pm^2|_{{\mathbb{R}}_{\sm_0}}
\otimes{\cal S}({\mathbb{R}}^3)\right)^\times
\end{equation}
realizations of the abstract rigged Hilbert spaces
\begin{equation}
\label{4.30b}
\tag{\ref{4.30}b}
\Phi_\pm\subset{\cal H}\subset\Phi_\pm^\times
\end{equation}
if the abstract spaces are algebraically and topologically (i.e.,
their meaning of convergence) equivalent to the corresponding function
spaces \cite{gelfand}. 

The energy wave function
$\<^-\sm|\psi^-\>\in\tilde{\cal S}\cap\H^2_+$ is a mathematical
realization of the vector $\psi^-\in\Phi_+$ and the vector $\psi^-$ is
a mathematical representation of the physical observable $\psi^-$
which is physically defined by the apparatus (detector) that registers
the observable, and similarly for $\phi^+\in\Phi_-$. 
Therefore, we have the following correspondences:
\begin{equation}
\nonumber
\text{registration apparatus}\ \Leftrightarrow\ \text{observable}\
|\psi^-\>\<\psi^-|\ \Leftrightarrow\ \text{wave function}\
\psi^-(\sm)=\<^-\sm|\psi^-\>
\end{equation}
\begin{equation}
\text{preparation apparatus}\ \Leftrightarrow\ \text{state vector}\
\phi^+\Leftrightarrow\ \text{wave function}\
\phi^+(\sm)=\<^+\sm|\phi^+\>\nonumber
\end{equation}
The modulus of the energy wave function $|\<^-\sm|\phi^+\>|^2$ describes the
energy distribution in the beam and the energy wave function
$|\<^-\sm|\psi^-\>|^2$ describes the energy resolution of the
detector.

We shall call all the function spaces that are intersections with the
spaces of Hardy class $\H_\pm^2|_{\mathbb{R}_{\sm_0}}$ Hardy
(function) spaces and we call  all (abstract) spaces $\Phi_\pm$
that have realization by Hardy function spaces also (abstract) Hardy
spaces.

In~\eqref{4.30} as in~\eqref{3.13} \eqref{3.14}, 
$\H_{+}^{2}$ means the functions of 
Hardy class analytic in the upper half of the second sheet
of the $\sm$-plane
and $\H_{-}^{2}$ means the functions of 
Hardy class analytic in its lower half.
Specifically, the physical values of  
$\<^{+}\sm-i0\,|\,\phi^{+}\>$ and of
$\<\psi^-|\sm-i0^-\>=\overline{\<^-\sm+i0|\psi^-\>}$ are the
boundary values of functions analytic in the lower half of the {\em second}
sheet~\footnote{if $f\in\H_+^2$, then its complex conjugate
function $\overline{f}\in\H_-^2$ and vice
versa. Therefore for the matrix elements $\<\psi^-|\phi^+\>$ we only require
the lower half of the complex energy plane (2nd sheet).}.
These
analyticity properties on the second sheet of the complex
$\sm$-Riemann surface will turn out to be important because the 
resonance poles of the $S$-matrix are located on the second 
Riemann sheet.

By virtue of Proposition~\ref{Property2.2}, 
the operator of total momentum $P_\mu=P_{a\mu}+P_{b\mu}$ and the invariant
mass square operator $M^{2}=P_\mu P^\mu$ are 
$\tau_{\Phi_{\pm}}$-continuous; 
hence their conjugates~\footnote{defined by 
 \eqref{r3} in Appendix~\ref{r}}, $P_{\mu}^{\times}$
and $M^{2^{\times}}$, are well defined on $\Phi_{\pm}^\times$~\footnote{To be precise, 
we should label $P_\mu$ etc by the space that
it acts on. $P_{\mu+}$ is the restriction to the space $\Phi_+$ of the
selfadjoint operator $\bar{P}_\mu$ in $\H$ and $P_{\mu-}$ is the
restriction to the space $\Phi_-$ of the selfadjoint operator
$\bar{P}_\mu$ in $\H$. $P_{\mu+}^\times$ is the conjugate operator of
$P_{\mu+}$ in $\Phi^\times_+$ which is a unique extension of
$P_\mu^\dagger=\bar{P}_\mu$ in $\H$ to $\Phi^\times_+$. 
$P_{\mu-}^\times$ is the conjugate operator of
$P_{\mu-}$ in $\Phi^\times_-$ which is a unique extension of  
$P_\mu^\dagger=\bar{P}_\mu$ in $\H$ to $\Phi_-^\times$.}.    
This can be seen by considering the realization, for instance, of the vectors
$P_\mu \psi^-$ and $M^{2}\psi^-$:
\begin{subequations}
\label{4.31}
\begin{align}
\label{4.31a}
\<P_\mu \psi^-|\cm\>&=\<\psi^-|P_\mu^\times|\cm\>
=\sqrt{\sm}\p_\mu\<\psi^-|\cm\>\,,\\
\label{4.31b}
\<M^2 \psi^-|\cm\>&=\<\psi^-|M^{2^\times}|\cm\>=\sm\<\psi^-|\cm\>\,.
\end{align}
\end{subequations}
According to Proposition~\ref{Property2.2} and the definition of the 
wave functions $\<\psi^-|\cm\>$ given in~\eqref{4.29a}, the multiplication 
operators by $\sqrt{\sm}\p_{\mu}$ and by $\sm$ which appear in 
the right hand side of~\eqref{4.31a} and~\eqref{4.31b} are
$\tau_{\Phi_+}$-continuous. Consequently, $P_\mu$ and $M^{2}$
are $\tau_{\Phi_+}$-continuous operators, and the conjugate
operators $M^{2^{\times}}$ and $P_{\mu}^{\times}$ that appear
in~\eqref{4.31} are everywhere defined, continuous
operators on $\Phi_+^\times$.  
Hence,~\eqref{4.31a} and~\eqref{4.31b} define, according to
\eqref{r6} \eqref{r7}, the functionals 
$|\cm\>$ as generalized eigenvectors of $P_\mu$ and 
$M^{2}$. The same discussion applies for the space $\Phi_-$. Summarizing
\begin{equation}
\label{4.32}
P_\mu\,:\,\Phi_{\pm}\rightarrow\Phi_{\pm}
\quad\text{ is }\ 
\tau_{\Phi_{\pm}}\text{-continuous}
\end{equation}
and
\begin{subequations}
\label{4.33}
\begin{equation}
\label{4.33a}
P_\mu^\times|\amp\>=\sqrt{\sm}\p_\mu |\amp\>\,,
\end{equation}
\begin{equation}
\label{4.33b}
M^{2^\times}|\amp\>=\sm|\amp\>\,.
\end{equation}
\end{subequations}
We can re-express the generalized eigenvalues of the momentum operator
in terms of the three velocity $\bs{v}$ by noting that
$\bs{\p}=\gamma\bs{v}=\frac{\bs{v}}{\sqrt{1-\bs{v}^{2}}}$,
and $1+\bs{\p}^{2}=\frac{1}{1-\bs{v}^{2}}=\gamma^{2}$.
Hence, the eigenvalues in~\eqref{4.33} can be rewritten
as
\begin{equation}
\label{4.34}
\begin{split}
H^\times|\amp\>&=\gamma\sqrt{\sm}|\amp\>\,,\\
\bs{P}^\times|\amp\>&=\gamma\sqrt{\sm}\bs{v}|\amp\>\,.
\end{split}
\end{equation}
Note that here $\sqrt{\sm}$ is not only restricted to the ``physical'' scattering
energies \eqref{val_a} but can be a complex value; in particular for the kets
$|\amr\>$, $\sm$ can be any value in the lower half complex plane (second sheet).

For the branch of $\sqrt{\sm}$ in~\eqref{4.31}, \eqref{4.33}
and \eqref{4.34}, we choose
\begin{equation}
\label{4.35}
-\pi\leq \rm{Arg}\,\sm<\pi\,.
\end{equation}
This choice of branch, even though irrelevant for the physical values
of $\sm$, will be needed since we will analytically continue
the kets $|\amp\>$ to the second Riemann sheet as described in
Section~\ref{sec5}.

We have now a well defined mathematical theory in which the momentum
and energy operators $P_\mu$ (and the other generators $J_{\mu\nu}$)
of the Poincar\'e group have a  well defined mathematical meaning. We denote the triplet whose
wave functions are given by the Hardy function spaces \eqref{4.29}
again by
\begin{eqnarray}
\Phi_-\subset&\H&\subset\Phi_-^\times\quad \text{for the prepared in-states}\
\phi^+\nonumber\\
\Phi_+\subset&\H&\subset\Phi_+^\times\quad \text{for the detected
out-observables}\
\psi^-\label{4.36}
\end{eqnarray}
In addition to the apparatus prepared states $\phi^+\in\Phi_-$ and the
apparatus detected observables $\psi^-\in\Phi_+$ one has the
generalized state vectors $F^\pm\in\Phi_\mp^\times$. An example of these are
the $F^\pm=|\bs{\p}j_3[\sm j]^\pm\>$. The standard interpretation of
these kets is as out-going and in-coming plane waves (in analogy 
to the non-relativistic case of Section~\ref{sec2}). The bra-kets
$|\<^+\sm|\phi^+\>|^2$ represent the energy distribution in the
prepared in-state, i.e., the probability density for the CM-energy
$\sqrt{\sm}$ and for the momentum $p=\sqrt{\sm}\p$ in the state
$\phi^+$, as prepared by the preparation apparatus (accelerator).
The bra-ket $|\<^-\sm|\psi^-\>|^2$ represents the energy resolution
of the detector $\psi^-$. The interpretation of the Born probability
$|(\psi^-,\phi^+)|$ is thus extended to the $|\<^+\sm|\phi^+\>|^2$
(probability density for the beam $\phi^+$, i.e., the probability
of the particles $a+b$ to have the energy
$\sqrt{\sm}$) and to the $|\<\psi^-|\sm^-\>|^2$ (probability for the
detector to register the particles $c+d$ with energy $\sqrt{\sm}$). 

If one uses the standard relationship \eqref{4.22} between
interaction-free and exact quantities, one can -- heuristically-- 
justify (as in Section \ref{sec3} for the non-relativistic case) that:
\begin{subequations}
\label{4.37}
\begin{equation}
|\<\phi^+|\sm^+\>|=|\<\phi^{in}|\sm\>|\label{4.37a}
\end{equation}
i.e., the energy distribution in the beam is measured in the asymptotic
interaction-free region.
Similarly
\begin{equation}
|\<\psi^-|\sm^-\>|=|\<\psi^{out}|s\>|\label{4.37b}
\end{equation}
\end{subequations}
i.e., the detector's energy resolution is the resolution of the asymptotic region.
Since we want to work in our mathematical theory only with the interaction
incorporating {\it exact} quantities, we only conclude from \eqref{4.37} that
$|\<\phi^+|\sm^+\>|^{2}$ and $|\<\psi^-|\sm^-\>|^2$ are as measured in the asymptotic
in and out region.

In addition to the eigenkets $|\bs{\p}j_3[\sm
j]^\pm\>\in\Phi_\mp^\times$ with real positive energy $\sqrt{\sm}$,
the spaces $\Phi_\mp^\times$ contain many other generalized vectors
$F^\pm$. In particular,
\begin{subequations}
\label{4.38}
\begin{equation}
\Phi^\times_-\ \text{contains eigenkets}\ |\sm^+\>\ \text{with complex
eigenvalue}\ \sm\ \text{of Im}\sm\geq0\label{4.38a}
\end{equation}  
\begin{equation}
\Phi^\times_+\ \text{contains eigenkets}\ |\sm^-\>\ \text{with complex
eigenvalue}\ \sm\ \text{of Im}\sm\leq0\label{4.38b}
\end{equation}
\end{subequations}
This is the important advantage that the functionals
$|\sm^\mp\>\in\Phi^\times_\pm$ of the Hardy spaces $\Phi_\pm$ have
over the ordinary Dirac kets which (if defined at all) are
defined as functionals over the Schwartz space and have therefore only
real generalized eigenvalues. 

As a consequence of \eqref{4.38}, the quantities $\<\psi^-|\sm^-\>$
and $\<\sm^+|\phi^+\>$ can be analytically continued from the real
positive energy axis into the {\em lower} half complex energy 
plane~\footnote{From $\<^-\sm|\psi^-\>\in\H_+^2$ follows that
$\<\psi^-|\sm^-\>\equiv\overline{\<^-\sm|\psi^-\>}\equiv\<^-\bar{\sm}|{\psi}^{-}\>\in\H_-^2$.}.
The kets $|\sm^\mp\>$ are thus more generalized vectors than the
ordinary Dirac kets because the $\<\psi^-|\sm^-\>$ and
$\<^+\sm|\phi^+\>$ are not only smooth rapidly decreasing functions of
$\sm$ but can also be analytically extended to complex $\sm$. This
analytic property of the $|\sm^\mp\>$ is already contained in its
infinitesimal form in the plane-wave solutions of the
Lippmann-Schwinger equations, where one describes the incoming and
outgoing boundary conditions by an infinitesimal complex energy
$\sm\pm i\epsilon$~\footnote{The
conventional Lippmann-Schwinger integral equation uses actually an
infinitesimal imaginary part of the non-Lorentz invariant energy
$p^0$, $\sm=(p^0\pm i\epsilon)^2-\bs{\p}^2=\sm\pm i2p^0\epsilon=\sm\pm
i\epsilon'$, which for infinitesimal $\epsilon$ is equivalent to using
an infinitesimal Lorentz invariant $i\epsilon$. We prefer to make the
analytic extension in the invariant energy squared $\sm$, since it is
the preferred variable used in the relativistic $S$-matrix.}. To go
from infinitesimal analyticity to the analyticity requirement  for the
whole semi-plane contained in the Hardy space hypothesis \eqref{4.36}
may appear a big jump. But since one cannot experimentally distinguish
between an energy resolution of a detector described by smooth
functions $|\<\sm|\psi\>|^2$ of energy squared $\sm$ and a smooth
$|\<^-\sm|\psi^-\>|^2$ that can be analytically continued into the
complex $\sm$-plane, the hypothesis is not in conflict with
observations. And, as we shall see below, the consequences of the 
hypothesis have many physical aspects which the Hilbert space axiom is
not capable of describing.

After we have established \eqref{4.37} in analogy to the usual heuristic
arguments for the non-relativistic case invoking
the Moeller wave operators $\Omega^\pm$, we know that
the energy distribution of the prepared (beam) state
$|\<^+\sm|\phi^+\>|^2$ and the energy resolution of the detector
$|\<^-\sm|\psi^-\>|^2$ is observed in the asymptotic in- and out-
regions, respectively. If we use this as postulate and define the
relativistic $S$-matrix (with 4-velocity basis) in analogy to \eqref{3.5f}
by
\begin{equation}
\<\bs{\p}j_3[\sm j]|S|\bs{\p}'j'_3[\sm j]\>\equiv\<^-\bs{\p}j_3[\sm
j]|\bs{\p}'j'_3[\sm' j']^+\>\label{4.39}
\end{equation}
we have no further need for the interaction free in-states
$\phi^{in}$, 
out-observables $\psi^{out}$ and basis vectors $|\bs{\p}'j'_3[\sm
j]\>$ nor for the 
interaction free Poincar\'e generators \cite{weinberg}
$P_{free}^\mu=\Omega_\mp^{-1}P^\mu\Omega_\mp,\
J_{free}^{\mu\nu}=\Omega_\mp^{-1}J^{\mu\nu}\Omega_\mp$. We can work entirely
with the states $\phi^+\in\Phi_-$, observables $\psi^-\in\Phi_+$, the
basis vectors $|\bs{\p}j_3[\sm j]^\pm\>\in\Phi_\mp^\times$ and the exact
generators $P_\mu$ and $J_{\mu\nu}$ of Poincar\'e transformations 
which incorporate the interactions
\cite{weinberg}. The Poincar\'e transformations $U_+(\Lambda,a)$ generated by $P_\mu,\
J_{\mu\nu}$,  in the space of observables $\Phi_+$ and the Poincare\'e
transformations
$U_-(\Lambda,a)$ in the space of in-states  $\Phi_-$ do not describe
kinematic translations and rotations like the interaction free
transformations but dynamical evolutions, e.g.,
$U_+(\Lambda,a)=e^{iP_0t}$ describes time evolution of the observable
$\psi^-$ (Heisenberg picture) and $U_-^\dagger(\Lambda,a)=e^{-iP_0t}$
describes the time evolution of the state $\phi^+$ (Schrodinger
picture) by the amount $t$ (in the rest frame of the state). The
property of the interaction is encapsulated in the (reduced)
$S$-matrix element.
For instance, the property that a resonance is formed in the reaction
\eqref{4.5} is described by a pole in the second sheet of the
$S$-matrix at the complex energy $\sm=\sm_R=(M-i\Gamma/2)^2$.  

In the following section we shall obtain the decaying state vectors,
the Gamow vectors, from the $S$-matrix pole. 
Our theoretical frame based on the new hypothesis \eqref{4.36}
allows us to describe
relativistic resonances as decaying states by vectors associated to
irreducible representations of Poincar\'e transformations,
this is similar to Wigner description of stable relativistic particles.
In a subsequent paper~\cite{rgv2}, 
we shall derive their transformation properties under Poincar\'e
transformations. 
The result will look similar to but will turn out to be very different 
from Wigner's Poincar\'e transformations, since it will distinguish 
the forward light cone and a direction of time. This transformation property will lead
to an unambiguous definition of resonance mass and width, which has
been an open problem \cite{tR98}.

\section{Derivation of the Relativistic Gamow Vectors from an
$S$-matrix Pole}\label{sec5}

To obtain the Gamow vector from the first order $S$-matrix pole~\footnote{
The same procedure can be used for higher order $S$-matrix poles which
leads to Jordan blocks of Gamow vectors, cf.~\cite{maxson}, for the non-relativistic
case}, we start from the $S$-matrix element $(\psi^-,\phi^+)$, use in it the
expansions \eqref{4.25}, \eqref{4.26+} and the relativistic analogy of the
definition \eqref{3.5g}~\footnote{This is just an other more familiar way of writing
the $S$-matrix elements on the right hand side.}:
\begin{equation}
\nonumber
\<\bs{\p}j_3[\sm j]\eta|S|\bs{\p'}j_3'[\sm'j']\eta'\>\equiv
\<^-\bs{\p}j_3[\sm j]\eta|\bs{\p'}j_3'[\sm' j']\eta'{}^+\>\,.
\end{equation}
Then we obtain
\begin{multline}
\label{5.1}
(\psi^{-},\phi^{+}) =\sum_{jj_{3}}\int \frac{d^{3}\hat{p}}{2\hat{E}}d\sm
\sum_{j'j_{3}'}\int \frac{d^{3}\hat{p}'}{2\hat{E}'}d\sm'
\langle\,\psi^{-}\,|\,\am\rangle\\
\langle\,\a\,|\,
S\,|\,\apr\,\rangle
\langle^{+}\apr\,|\,\phi^{+}\,\rangle
\end{multline}
\begin{equation}
\nonumber
\text{where } \hat{E}=\hat{E}(\bs{\p}) \equiv \sqrt{1+\bs{\p}^2} = \p^0\,.
\end{equation}
\begin{comment}
where we insert into $(\psi^{-},\phi^{+})$ complete systems of 
basis vectors~\footnote{
We ignore the possible existence of bound states of $H$ of which there
are usually none; certainly not for the $\pi^{+}\pi^{-}$ system of 
$\pi^{+}\pi^{-}\rightarrow \rho^{0}\rightarrow \pi^{+}\pi^{-}$.}
and use the expansions \eqref{4.25} and \eqref{4.26+}.  
\end{comment}
Using the invariance of the $S$ operator with respect to space 
time translations one can show \cite{weinberg} that the $S$-matrix can be written in
the following way
\begin{eqnarray}
\nonumber
\lefteqn{\langle\,\a\eta\,|\,S\,|\,\apr\eta'\,\rangle}\\
\label{5.2}
& &\qquad\qquad =\delta(\bs{p}-\bs{p'})\delta(p_{0}-p_{0}')
\langle\!\langle\,\a\eta\,|\,\tilde S\,|\,\apr\eta'\,\rangle\!\rangle
\end{eqnarray}
where $\langle\!\langle\quad|\tilde S|\quad\rangle\!\rangle$ is a 
reduced $S$--matrix element defined by \eqref{5.2}. 
For later considerations it is important to note that the invariance
does not have to be considered for the whole group of transformations;
it is sufficient to consider a subsemigroup to obtain \eqref{5.2}.
The equation \eqref{5.2} can also be written
\begin{multline}
\label{5.3}              
\langle\,\a\eta\,|\,S\,|\,\apr\eta'\,\rangle
=2\hat{E}(\hat{p})\delta(\bs{\hat{p}}-\bs{\hat{p}'})\delta(\sm-\sm')\\
\langle\!\langle\,\a\eta\,|\,S\,|\,\aprr\eta'
\,\rangle\!\rangle
\end{multline}
where $\<\!\<\quad|\,S\,|\quad\>\!\>$ is another reduced 
matrix element defined by \eqref{5.3}.
In \eqref{5.2} and
\eqref{5.3} we include explicitly the degeneracy
quantum numbers and the species labels which we denote collectively by
$\eta$ for purposes of clarity and completion, 
but we will omit it below for the sake
of notational convenience. 
The form \eqref{5.3} follows from \eqref{5.2}
by the defining identities
$\bs{\p}=\frac{\bs{p}}{\sqrt{\sm}}$, 
$\hat{p}^{0}=\frac{p^{0}}{\sqrt{\sm}}$.

We now use the invariance of the $S$-matrix with respect to Lorentz
transformations either in the form \cite{weinberg}
\begin{subequations}
\label{5.4}
\begin{equation}
\label{5.4a}
\left(U(\Lambda)\psi^-,U(\Lambda)\phi^+\right)=(\psi^-,\phi^+)\,,
\end{equation}
or in the form \cite{weinberg}
\begin{equation}
\label{5.4b}
U^\dagger_{\rm free}(\Lambda)SU_{\rm free}(\Lambda)=S\,.
\end{equation}
\end{subequations}
With this we further simplify \eqref{5.3}. We first choose $\Lambda=L^{-1}(\p)$
where $L(\p)$ is the boost (rotation free Lorentz transformation)
\begin{equation}
\label{trans}
L^\kappa_{\hphantom{\mu}\nu}=
\left(
\begin{array}{cc}
\frac{p^0}{m}&-\frac{p_n}{m}\\
\frac{p^\kappa}{m}&\delta^\kappa_n\!-\!\frac{\frac{p^\kappa}{m}
\frac{p_n}{m}}{1+\frac{p^0}{m}}
\end{array}
\right)\,.
\end{equation}
$L(\p)$ depends upon the parameter $\p=\frac{p}{m}\in{\mathbb{R}}$ and has the
property
\begin{equation}
\label{prop}
L^{-1}(\p)^\mu_{\,\nu}p^\nu=\left(\begin{array}{c}m\\0\\0\\0\end{array}\right)
\end{equation}
From \eqref{5.4b} follows
\begin{eqnarray}
\<\<\bs{\p}j_3[\sm
j]|U^\dagger(L^{-1}(\p))SU(L^{-1}(\p))|\bs{\p}j'_3[\sm j']\>\>
&=&
\<\<\bs{0}j_3[\sm j]|S|\bs{0}j'_3[\sm j']\>\>\nonumber\\
&\equiv&\<\<j_3[\sm j]|S|j'_3[\sm j]\>\>\label{reduced_element}
\end{eqnarray}
for all $\bs{\p}\in{\mathbb{R}^3}$. This means the reduced $S$-matrix element is 
the same for 
all $\hat{p}$ as in the center of mass frame, i.e., for
$\bs{\p}=\bs{0}$. 

Invariance with respect to rotations, $\Lambda = {\cal R}$ in the 
center of mass frame shows then by analogous arguments for the discrete 
quantum numbers
$j_{3}$ and $j$ that the reduced matrix element is proportional to
$\delta_{j_{3}j_{3}'}\delta_{jj'}$ and independent of $j_{3}$.
Since Poincar\'e transformations do not change the 
Poincar\'e invariants $\sm$ 
and $j$, the reduced matrix element can still depend upon $\sm$
and $j$. Thus we have
\begin{multline}
\label{red_mat}
\langle\,\a\eta\,|\,S\,|\,\apr\eta'\,\rangle
=2\hat{E}(\hat{p})\delta(\bs{\hat{p}}-\bs{\hat{p}'})\delta(\sm-\sm')
\delta_{j_{3}j_{3}'}\delta_{jj'}\\
\langle\,\eta\,\|\,
S_{j}(\sm)\,\|\,\eta'\,\rangle
\end{multline}
If there are no degeneracy quantum numbers, or if we suppress the 
particle species label and channel numbers and restrict ourselves to the
case without spins (like for the $\pi^{+}\pi^{-}$ system), then the
reduced matrix element can be written as 
\begin{equation}
\label{5.5}
\langle\,\eta\,\|\,S_{j}(\sm)\,\|\,\eta'\,\rangle=S_{j}(\sm)=\left\{\begin{array}{c}
2ia_j(\sm)+1\ \text{for elastic scattering}\ \eta=\eta'\\
2ia^{(\eta)}_j(\sm)\ \text{for a reaction from state $\eta'$ to $\eta$}\end{array}\right.
\end{equation}
where $j$ is the total angular momentum in the center of mass, and 
$a_j(\sm)$ is the $j$-th partial wave amplitude for elastic scattering
and $a^\eta_j(\sm)$ is the amplitude for inelastic scattering (from $\eta'$ into
channel $\eta$). We insert 
\eqref{red_mat} (or \eqref{5.5}) into
\eqref{5.1} and integrate over $\bs{\p}$ and $\sm$ 
to obtain for the $S$-matrix element
\begin{equation}
\label{5.6}
(\psi^{-},\phi^{+})=\sum_{j}\int_{\sm_0}^{\infty}d\sm
\sum_{j_{3}}\int\frac{d^{3}\hat{p}}{2\p^0}
\<\psi^-|\bs{\p}j_3[\sm j]^-\>S_j(\sm)\<^+\bs{\p}j_3[\sm j]|\phi^+\>
\end{equation}
\begin{comment}
Since resonances come in one partial wave with definite
value of angular momentum, 
\end{comment}

Resonances are usually associated with a fixed value of $j$ or ($j$, parity).
This means, in the partial wave analysis of the experimental (differential 
cross section) data, a resonance is identified by one (or a superposition of
several) Breit-Wigner amplitudes \eqref{sdescription} in a partial wave
amplitude $a_j(\sm)$. The value $j$ for this partial wave amplitude is then reported
as the spin (or $j^P$) of the resonance \cite{particledata}. In the $S$-matrix
theory, resonances are defined as poles of the $j$-th partial $S$-matrix, $S_j(\sm)$,
at a complex value $\sm_R$. Thus, by todays theoretical definition and experimental 
analysis, a resonance is assigned to one partial wave with definite value of angular
momentum $j$. Therefore, we consider of \eqref{5.6} only the 
term with the resonating partial wave $j=j_{R}$ (e.g., 
$j_R=1$ for $\pi^{+}\pi^{-}\rightarrow \rho^{0} \rightarrow \pi^{+}\pi^{-})$), 
i.e., we restrict ourselves to the subspace
with $j=j_{R}\,(s=0,\,l=j,\, n=n_{\rho},\,n_{\pi\pi})$. 
This means that we consider only the term with $j=j_{R}$ in the sum
on the right hand side of \eqref{5.6} and call $S_{j_{R}}(\sm)=S_j(\sm)$.

We can further simplify our notation. After we employed the 4-velocity basis
vectors $|\bs{\p}j_3[\sm j]\eta^\mp\>$ to make use of Poincar\'e
invariance, we can ignore the quantum numbers $j,\ \eta$ because they
are fixed, and we can also suppress the quantum numbers $\bs{\p},\ j_3$ because they are summed
(integrated) over in \eqref{5.6}. We therefore use again the notation \eqref{4.25}
\begin{eqnarray}
\<^-\bs{\p}j_3[\sm
j]|\psi^-\>\rightarrow\<^-\sm|\psi^-\>&\in&\tilde{\cal
S}\cap\H^2_+|_{\mathbb{R}_{\sm_0}}\label{5.7}\\
\<^+\bs{\p}j_3[\sm
j]|\phi^+\>\rightarrow\<^+\sm|\phi^+\>&\in&\tilde{\cal
S}\cap\H^2_-|_{\mathbb{R}_{\sm_0}}\label{5.8}
\end{eqnarray}
and also write for the basis vectors the short form \eqref{4.25d}
\begin{equation}
\label{5.9}
|\bs{\p}[\sm j]^\mp\>\rightarrow|\sm^\mp\>\in\Phi^\times_\pm
\end{equation}
We assume now that our following analytic extension in the variable
$\sm$ does not effect the values of the quantum numbers $j_3$ and
$\bs{\p}=\frac{\bs{p}}{\sqrt{\sm}}$ which will be justified below. Then we can write 
the $j$-th partial $S$-matrix element of \eqref{5.6} as
\begin{eqnarray}
(\psi^-,\phi^+)_j&=&\int_{\sm_0}^{\infty}
d\sm\sum_{j_3}\int\frac{d^3\p}{2\p^0}\<\psi^-|\bs{\p}j_3[\sm
j]^-\>S_j(\sm)\<^+\bs{\p}j_3[\sm j]|\phi^+\>\nonumber\\
&\equiv&\int_{\sm_0}^\infty
d\sm\<\psi^-|\sm^-\>S_j(\sm)\<^+\sm|\phi^+\>
\label{5.10}
\end{eqnarray}
According to the standard analyticity assumptions \cite{eden} of
the $j$-th partial $S$-matrix, $S_j(\sm)$ is an analytic function on
the first (``physical'') sheet. The boundary values of this analytic
function on the real axis $\sm+i\epsilon,\ \epsilon\rightarrow0$ are
the  ``physical'' values that appear in the integral
\eqref{5.10}. There may be poles on the real axis for values
$\sm<\sm_0$, i.e., below the elastic scattering threshold
$\sm_0=(m_a+m_b)^2$. We ignore here such stable, bound state
poles, if they exist. 
The two sheeted Riemann surface (in the simplest case that we consider here) 
has a cut that starts at $\sm=\sm_0$. To reach the
second sheet one burrows through
the cut, Figure~\ref{contour_fig}. The integration contour
of the integral \eqref{5.10} extends along the lower
edge of the first sheet, right above the cut. 
If there is no further cut, which is the case we want
to consider for the time being, then
$S(\sm+i\epsilon)=S(\sm-i\epsilon_{II})$ along the cut $\sm,\
\sm_0\leq\sm<\infty$ where $\sm-i\epsilon_{II}$ is on the second
sheet. Thus we can as well extend the
integration along the upper edge of the second sheet just below the cut.
The second sheet of $S(\sm)$ can contain
singularities. We want to consider the case that there are only pole-singularities of the $S$-matrix. 

In particular, we shall consider only first order
poles and for the sake of definiteness we shall assume the case that
there are 2 (or $N$) first order poles. 
Poles of higher order can be treated in a similar way and lead to
Gamow states described by non diagonalizable density operators (and
Jordan blocks) \cite{maxson}. Cut-singularities in the lower half-plane
can also be
accommodated by additional background integrals, which we also do not want to consider here.

The $S$-matrix definition of a resonance is a
 first order pole on the second Riemann sheet at
$\sm_R=(M_R-i\Gamma_R/2)^2$. This definition is of practical importance only for values
of $\frac{\Gamma_R}{M_R}\leq10^{-1}$. Unstable states with
$\frac{\Gamma_R}{M_R}\approx10^{-3}-10^{-1}$ are usually called
relativistic resonances, while those with
$\frac{\Gamma_R}{M_R}\approx10^{-8}-10^{-16}$ are called decaying
relativistic particles (cf.~Section \ref{sec1}).

The particular parameterization of the complex pole position $\sm_R$ in
terms of $M_R$ and $\Gamma_R$ is still arbitrary and will be given a
physical meaning by our subsequent considerations.

Hermitian analyticity (symmetry relation of the $S$--matrix 
$S(\sm-i\epsilon)=S^{*}(\sm+i\epsilon)$) implies that when there is a 
pole $P$ at the complex position $\sm_R$ then 
there must also be a pole $P'$ at the complex conjugate position
$\sm_{R}^{*}=(M_{R}+\frac{i}{2}\Gamma_{R})^{2}$ on the second sheet
reached by burrowing through the cut from the lower half plane
of the physical sheet.
Thus a scattering resonance is defined by a pair of poles on the second 
sheet of the analytically continued $S$-matrix located at positions
that are complex conjugates of each other.
The pole $P'$ corresponds to the time-reversed situation which we
do not want to discuss here.
There may exist other resonance poles located at the same or
other physical sheets,
but we will restrict ourselves here to $N=2$  poles in the 
lower half plane second sheet at $\sm_{R_1}$ and $\sm_{R_{2'}}$ as shown
in Figure~\ref{contour_fig}. 

The contour of integration parallel to the real axis in the second sheet
over $\sm_0<\sm<\infty$ in \eqref{5.10} can now be deformed into the contour  shown
in Figure~\ref{contour_fig}: an integral from $\sm_0$ to $-\infty_{II}$, then
along the infinite semicircle ${C}_\infty$, around the two
poles ${C}_1$, ${C}_2$ and again along the infinite semicircle
${C}_\infty$. After this contour deformation,
\begin{comment}
In order to write the integral in \eqref{5.10} in terms of integrals
around the pole and something else, we deform the contour of
integration in \eqref{5.10} from the real line just above the real
axis on the first sheet $\sm+i\epsilon$ into path in the second sheet
by going through the cut as shown in Figure~\ref{contour_fig}.
\end{comment}
the integral in \eqref{5.10}  becomes (dropping the $j$ notation):
\begin{eqnarray}
(\psi^-,\phi^+)&=&\int_{\sm_{0}}^{-\infty_{II}}d\sm\<\psi^-|\sm^-\>S_{II}(\sm)\<^+|\phi^+\>\nonumber\\
&&+\int_{C_1}d\sm\<\psi^-|\sm^-\>S_{II}(\sm)\<^+\sm|\phi^+\>\nonumber\\
&&+\int_{C_2}d\sm\<\psi^-|\sm^-\>S_{II}(\sm)\<^+\sm|\phi^+\>\nonumber\\
&&+\int_{C_\infty}d\sm\<\psi^-|\sm^-\>S_{II}(\sm)\<^+\sm|\phi^+\>\label{5.11}
\end{eqnarray}
where $C_i$ is the circle (in the negative direction) around the pole at $\sm_{R_i}$. The
first integral in \eqref{5.11} extends along the negative real axis in the second sheet
(indicated by $-\infty_{II}$). The fourth integral is along the
infinite semicircle $C_{\infty}$. 

In order for this path deformation to be possible, the integrand in
\eqref{5.10} and \eqref{5.11} must be well defined in the area into
which the path will be deformed. This will be in our case the whole
real line at the upper edge of the second sheet, i.e., for 
\begin{equation}
\<\psi^-|\sm_{II}-i\epsilon^-\>S_{II}(\sm_{II}-i\epsilon)
\<^+\sm_{II}-i\epsilon|\phi^+\>,\quad
-\infty<\sm_{II}<\infty\nonumber
\end{equation}
and the entire lower plane of the second sheet. That $S(\sm)$ is well
defined on the whole Riemann surface except for the singularities
discussed above, and that it is bounded by a polynomial, i.e., that
there is a polynomial $P(\sm)$ such that 
\begin{equation}
\label{5.12}
|S_{II}(\sm)|\leq|P(\sm)|\quad \text{for large}\ |\sm|\,,
\end{equation}
are the standard assumptions of the $S$-matrix theory \cite{eden}.

The functions $\<\psi^-|\sm^-\>$ and $\<^+\sm|\phi^+\>$ are known for
the physical values $\sm=\sm_I+i\epsilon=\sm_{II}-i\epsilon,\
{\mathbb{R}}_{\sm_0}=\{\sm:\ \sm_0\leq\sm_{II}<\infty\}$.  
Our new hypothesis \eqref{4.29}, \eqref{4.36} tells us that they are
Hardy functions on ${\mathbb{R}_{\sm_0}}$, and since we are concerned
with the second sheet for resonances, \eqref{4.29} or \eqref{5.7} \eqref{5.8} must
precisely mean Hardy with respect to the second sheet. 
From the van-Winter theorem (\eqref{h3}, Appendix~\ref{h}) it follows that every
Hardy function is completely determined from its values on a half axis
of the real line. In other words, there exists a bijective mapping
\begin{equation}
\theta:\ \tilde{\cal S}\cap\H^2_\mp\rightarrow \left. \left( \tilde{\cal
S}\cap\H^2_\mp \right) \right|_{\mathbb{R}_{\sm_0}}\label{5.13}
\end{equation}
This means that the Hardy functions on the negative real axis
$-\infty<\sm_{II}<\sm_0$ are already completely determined from their
values for $\sm\in{\mathbb{R}_{\sm_0}}$ (scattering energies). Thus the
$\<^-\psi|\sm^-\>$ and $\<^+\sm|\phi^+\>$ are known for the entire
real axis, second sheet. From this they can be determined on the
entire semiplane using Titchmarsh theorem (\eqref{h:1}, Appendix~\ref{h}).  
As a consequence of these remarkable properties of the Hardy	
functions, the integrand of \eqref{5.10} is uniquely defined on
the whole lower semiplane second sheet and we can do the contour deformation throughout
the lower semiplane, second sheet, of the complex energy Riemann
surface.

Further, the integral along $C_\infty$ in \eqref{5.11} vanishes. To see
this, notice that from \eqref{5.12}
\begin{equation}
\int_{C_\infty}|d\sm\<\psi^-|\sm^-\>S(\sm)\<^+\sm|\phi^+\>
\leq\int_{C_\infty}|d\sm\<\psi^-|\sm^-\>P(\sm)\<^+\sm|\phi^+\>|\label{5.14}
\end{equation}
From Proposition \ref{Property2.2}, it follows that
$P(\sm)\<^+\sm|\phi^+\>\in\tilde{\cal S}\cap\H_-^2$. Hence, a
straightforward application of H\"older's inequality shows that
\begin{equation}
\<\psi^-|\sm^-\>P(\sm)\<^+\sm|\phi^+\>\in\H_-^1\label{5.15}
\end{equation}
With \eqref{5.14} and \eqref{5.15}, the vanishing of the integral on
$C_\infty$ follows then from Corollary B.1, Appendix B.

We shall now consider each of the remaining integrals in
\eqref{5.11}. The first integral has nothing to do with any of the
resonances; it is the non-resonant background term
\begin{equation}
\int_{\sm_0}^\infty
d\sm\<\psi^-|\sm^-\>S_{II}(\sm)\<^+\sm|\phi^+\>\equiv\<\psi^-|\phi^{\rm
bg}\>\label{5.16}
\end{equation}
which we express as the matrix element of $\psi^-$ with a generalized
vector $\phi^{\rm bg}$ that is defined by it. It will not be further
discussed in the present section.

In the integrals along the circles $C_i$ around the poles $\sm_{R_i}$
we use the expansion
\begin{equation}
S(\sm)=\frac{R^{(i)}}{\sm-\sm_{R_i}}+R_0+R_1(\sm-\sm_{R_i})+\cdots\label{5.17}
\end{equation}
for each of the two (or $N$) integrals {\em separately}. The integrals
around the poles, the pole terms, are calculated in the following way
\begin{eqnarray}
(\psi^-,\phi^+)_{\text{pole
term}}&=&\int_{C_i}d\sm\<\psi^-|\sm^-\>S(\sm)\<^+\sm|\phi^+\>\label{5.18}\\
&=&\int_{C_i}d\sm\<\psi^-|\sm^-\>\frac{R^{(i)}}{\sm-\sm_{R_i}}\<^+\sm|\phi^+\>\label{5.19}\\
&=&-2\pi
iR^{(i)}\<\psi^-|\sm_{R_i}^-\>\<^+\sm_{R_i}|\phi^+\>\label{5.20}\\
&=&\int_{-\infty_{II}}^\infty
d\sm\<\psi^-|\sm^-\>\frac{R^{(i)}}{\sm-\sm_{R_i}}\<^+\sm|\phi^+\>\label{5.21}
\end{eqnarray}
To get from \eqref{5.19} to \eqref{5.20}, the Cauchy theorem has
been 
applied. To get from \eqref{5.19} to \eqref{5.21}, the contour $C_i$
of each integral has been separately deformed into the integral along
the real axis from $-\infty_{II}<\sm<\infty$ (and an integral along
the infinite semicircle, which vanishes because of the Hardy class
property). The equality \eqref{5.20} and \eqref{5.21} is the
Titchmarsh theorem for Hardy class functions (B.1, Appendix B).

The integral \eqref{5.21} extends from $\sm=-\infty$ in the second
sheet along the real axis to $\sm=\sm_0$ and then from $\sm=\sm_0$ to
$\sm=\infty$ in either sheet. (It does not matter whether we take the
second part of the integral over the physical values $\sm,\
\sm_0\leq\sm<\infty$ immediately below the real axis in the second
sheet or in the first sheet immediately above the real axis).  The
major contribution to the integral comes from the physical values
$\sm_0\leq\sm<\infty$, if the pole position $\sm_{R_i}$ is not too far
from the real axis. The integral in \eqref{5.21} contains the
Breit-Wigner amplitude
\begin{equation}
a_j^{BW_i}=\frac{R^{(i)}}{\sm-\sm_{R_i}},\ \text{but with}\
-\infty_{II}<\sm<\infty\label{5.22}
\end{equation}
Unlike the conventional Breit-Wigner for which $\sm$ is taken over
$\sm_0\leq\sm<\infty$, the Breit-Wigner in \eqref{5.22} is an
idealized or exact Breit-Wigner whose domain extends to $-\infty_{II}$
in the second (unphysical) sheet.

By \eqref{5.21} we have associated each resonance pole at $\sm_{R_i}$ to
an exact Breit-Wigner \eqref{5.22} which we obtain by omitting the
integral over the well behaved functions
$\overline{\<^-\sm|\psi^-\>}\<^+\sm|\phi^+\>\in\tilde{\cal
S}\cap\H_-^2$ from \eqref{5.21}. By \eqref{5.20} we have associated
each resonance pole at $\sm_{R_i}^-$ with vectors
$|\sm_{R_i}^-\>=|\bs{\p}j_3[\sm_{R_i}j]^-\>$ which we call 
in analogy to \eqref{bwgk} the relativistic Gamow vector or Gamow 
ket~\footnote{The term relativistic Gamow vector has been used before in
an other attempt to extend \eqref{bwgk} into the relativistic domain \cite{pronko};
it is not clear to what extent these are related to \eqref{defining_rel}.
The relativistic Gamow vectors of the Poincar\'e semigroup \eqref{defining_rel}
\eqref{5.23} were introduced in \cite{kielanowski}.}.

We obtain a representation of the Gamow vector by using
the equality \eqref{5.20}=\eqref{5.21} and omitting the arbitrary
$\psi^-\in\Phi_+$ (which represents the decay products defined by the
detector). Thus  the defining relation of this ket (functional over $\Phi^+$)
is
\begin{equation}
\label{defining_rel}
|\sm_{R_i}^-\>=\frac{i}{2\pi}\int_{-\infty_{II}}^{+\infty}|\sm^-\>
\frac{1}{\sm-\sm_{R_i}}\frac{\<^+\sm|\phi^+\>}{\<^+\sm_{R_i}|\phi^+\>}\,.
\end{equation}
But various other ``normalizations'' will also be used, e.g., the one
with the factor $\sqrt{2\pi\Gamma}$ in \eqref{bwgk}. Reverting the short
form notation \eqref{5.9} and using
the notation that includes the degeneracy quantum numbers,
 the relativistic Gamow kets are usually defined as:
\begin{equation}
|\bs{\p}j_3[\sm_{R_i}j]\eta^-\>=\frac{i}{2\pi}\int_{-\infty}^\infty
 d\sm|\bs{\p}j_3[\sm
 j]\eta^-\>\frac{1}{\sm-\sm_{R_i}}\label{5.23}
\end{equation}

The Gamow kets \eqref{5.23} are a superposition of the exact --not
asymptotically free -- ``out-plane waves'' $|\bs{\p}j_3[\sm
j]\eta^-\>$. The degeneracy quantum numbers $\eta$ of the Gamow kets
$|\bs{\p}j_3[\sm_R j],\eta^-\>$ are the same as the ones chosen for the
Lippmann-Schwinger out plane wave kets $|\bs{\p}j_3[\sm
j]\eta^-\>$. However, whereas for the Lippmann-Schwinger kets one can
choose generalized eigenvectors of any complete set of commuting
observables, e.g., one could choose momentum eigenkets
$|\bs{p}j_3[\sm j]\eta^-\>$, one does not have the same freedom for
the Gamow kets. Since in the contour deformations that one uses to get
from \eqref{5.10} to \eqref{5.11}, and ultimately to
\eqref{5.18}--\eqref{5.21}, one makes an analytic extension in the
variable $\sm$ to complex values. If one chooses the momentum to label
the basis vectors then because of $p_\mu p^\mu=\sm$, the $p_\mu$
also change and become complex when $\sm$ is extended to the complex
plane. Thus, $p_\mu$ could not be kept at one and the same value
during this analytic continuation  and the Gamow vector on the
l.h.s. of \eqref{5.23} would be a complicated (continuous)
superposition (integral) over different values of $\bs{p}$ and not
just a superposition over different values of $\sm$. For this reason,
the momentum $\bs{p}$ is not a good choice as a label for the basis
vectors. In contrast, the space components of the 4-velocity
$\bs{\p}=\bs{p}/\sqrt{\sm}$ is a good choice because then we can
impose the condition that $\bs{p}$ will become complex in the analytic
continuation to complex $\sm$ in such a way that $\p^\mu=p^\mu/\sqrt{\sm}$ will always
be real. This condition restricts the arbitrariness of analytic
continuation and makes the momentum only ``minimally complex''. 
As we shall discuss later, minimally complex momentum keeps the
representations of the Lorentz subgroup of the Poincar\'e group
unitary. In the analytic continuation in $\sm$ under the restriction
that $\p$ be real, only the representations of the space-time
translations turn into (causal) semigroup representations. The
homogeneous Lorentz transformations are the same as in Wigner's
representations. We will call this subclass of semigroup
representations of $\P$ minimally complex. They will be the
subject of a subsequent paper \cite{rgv2}.

With \eqref{5.22} and \eqref{5.23}, we have obtained for each
resonance defined by the pole of the $j$-th partial $S$-matrix at
$\sm=\sm_{R}$ an ``exact'' Breit-Wigner \eqref{5.22} and associated
to it a set of ``exact'' Gamow kets \eqref{5.23}. These Gamow kets
\eqref{5.23} span, like the Dirac kets $|\bs{p}j_3[\sm j]\eta\>$
\eqref{4.1}, \eqref{4.11} over the Schwartz space $\Phi$, the space of
an irreducible representation $[\sm_R j]$ of Poincar\'e transformations.
But, unlike the space spanned by the ordinary Dirac kets, the
representation space spanned by the kets of \eqref{5.23} is not a
representation space of a unitary group transformations.

We have the correspondence
\begin{eqnarray}
\!\!\!\text{Exact Breit-Wigner}&\Leftrightarrow&\text{Exact Gamow
Vector}\nonumber\\
\!\!\! a_j^{BW_i}(\sm)=\frac{R_i}{\sm-\sm_R}&\Leftrightarrow&|[\sm_R
j]f^-\>=\sum_{j_3}\int\frac{d^3{\p}}{2\p^0}|\bs{\p}j_3[\sm_R
j]^-\>f_{j_3}(\bs{\p})\label{5.24}\\
\!\!\! \text{for}\ -\infty_{II}<\sm<\infty&&\text{for
all functions}\ f_{j_3}(\bs{\p})\in{\cal S}({\mathbb{R}}^3),\ -j\leq j_3\leq j\nonumber
\end{eqnarray} 
The Gamow vectors $|\bs{\p}[\sm_R j]f^-\>$ have, according to
\eqref{5.22} and \eqref{5.23}, also the representation
\begin{equation}
|[\sm_Rj]f^-\>=\frac{i}{2\pi}\int_{-\infty_{II}}^\infty d\sm|[\sm
 j]f^-\>\frac{1}{\sm-\sm_R}\label{5.25}
\end{equation}
They are functionals on the Hardy space $\Phi_+$, i.e.,
$|[\sm_Rj]f^-\>\in\Phi_+^\times$.

Equation \eqref{5.25} is reminiscent of the continuous basis vector
expansion \eqref{4.26-} of $\psi^-\in\Phi_+\subset\H$ with respect to
the generalized eigenvectors $|[\sm j]f^-\>$ of $P_\mu P^\mu$ with
eigenvalue $\sm$. However, in \eqref{4.26-} $\sm$ extends over
$\sm_0\leq\sm<\infty$, whereas in \eqref{5.25} $\sm$ extends
over $-\infty_{II}<\sm<+\infty$ and the ``wave function''
$\psi^G(\sm)\equiv\frac{i}{2\pi}\frac{1}{\sm-\sm_R}$ is not a 
well behaved Hardy  function like $\psi^-(\sm)\equiv\<^-\sm|\psi^-\>\in{\cal
S}\cap\H_+^2|_{\mathbb{R}_{\sm_0}}$ of \eqref{4.26-}. Also, in the exact Breit-Wigner ``wave
function'' $\psi^G(\sm)=\frac{i}{2\pi}\frac{1}{\sm-\sm_R}$ in \eqref{5.25}, the variable $\sm$ extends
over $-\infty<\sm<\infty$. Thus the continuous linear superpositions
\eqref{5.25}, which define the relativistic Gamow vectors, are
entirely different mathematical entities than the $\psi^-$ of
\eqref{4.26+}. The Gamow vectors $|[\sm_R j]f ^- \>$ and
also the Gamow kets  $|\bs{\p}j_3[\sm_Rj]\eta^-\>$ of \eqref{5.23}
are functionals over the space $\Phi_+$. 
The equations \eqref{5.25} and \eqref{5.23} are thus 
functional equations over
the space  $\Phi _+$, and \eqref{5.23} can be stated in terms of
the smooth Hardy  class functions  $\overline{\psi}^-({\mathsf
s})\equiv\overline{\<^-[\sm j]|\psi^-\>}=\<\psi ^-|\bs{\p}j_3[\sm j]\eta^-\>\in \tilde{\cal S}
\cap {\cal H}^2_-$ as
\begin{eqnarray}
   \langle \psi^-|\bs{\p}j_3[\sm_Rj]\eta^-\>
       &\equiv& -\frac{i}{2\pi} \oint d{\mathsf s}\,
   \langle \psi^-|\bs{\p}j_3[\sm j]\eta ^->
   \frac{1}{{\mathsf s}-{\mathsf s}_R}\text{~\footnote{}}
   \label{5.26} \\
   \langle \psi^-|\bs{\p}j_3[\sm_Rj]\eta^-\>
   &=&\frac{i}{2\pi} \int_{-\infty_{II}}^{+\infty} 
   d{\mathsf s}\, \langle \psi^-|\bs{\p}j_3[\sm j]\eta^-\>
   \frac{1}{\mathsf{s}-\mathsf{s}_R}\label{5.27}
\end{eqnarray}
\begin{comment}
\begin{subequations}
\label{5.26-}
\begin{equation}
   \langle \psi^-|\bs{\p}j_3[\sm_Rj]\eta^-\>
       \equiv -\frac{i}{2\pi} \oint d{\mathsf s}\,
         \langle \psi^-|\bs{\p}j_3[\sm j]\eta ^->
        \frac{1}{{\mathsf s}-{\mathsf s}_R}
        \label{5.26}\tag{\ref{5.26-}~\footnote{}}
\end{equation}
\end{subequations}
\begin{equation}
  \langle \psi^-|\bs{\p}j_3[\sm_Rj]\eta^-\>
  =\frac{i}{2\pi} \int_{-\infty_{II}}^{+\infty} 
  d{\mathsf s}\, \langle \psi^-|\bs{\p}j_3[\sm j]\eta^-\>
  \frac{1}{\mathsf{s}-\mathsf{s}_R}\label{5.27}
\end{equation}
\end{comment}
\footnotetext{The integral $\oint$ is defined to be counter-clockwise whereas the integration
around $C_i$ in \eqref{5.19} is clockwise, cf.~Figure~\ref{contour_fig}. This explains the sign
difference.}for all $\psi^-\in\Phi_+$. This is just another form of the definition
\eqref{5.25} in terms of the well-behaved functions $\<\psi^-|\bs{\p}j_3[\sm j]\eta^-\>$
rather than the singular kets $|\bs{\p}j_3[\sm j]\eta^-\>$; \eqref{5.27} is the 
Titchmarsh theorem for the Hardy function $\<\psi^-|\bs{\p}j_3[\sm j]\eta^-\>$.

The first equality \eqref{5.26} 
is again the well known Cauchy formula for the 
analytic function 
$\overline{\psi^-} ({\mathsf s})=\langle \psi ^-|{\mathsf s}^-\rangle$.  
The second 
equality \eqref{5.27} is the Titchmarsh theorem (B1. of Appendix B) 
for the Hardy class function 
$\overline{\psi^-}({\mathsf s})$ in the lower half plane of the second 
sheet.  
The integration path extends as in \eqref{5.25} along the 
real axis in the second sheet, which agrees for physical values 
$\sm_0\leq {\mathsf s}<\infty$ only with the integration along the 
real axis on the first sheet.

With \eqref{5.24} we have associated a space of vectors \eqref{5.25}
with the Breit-Wigner partial wave amplitude \eqref{5.22}. The
relativistic Breit-Wigner \eqref{5.22} is the pole term 
of the relativistic $j$-th partial $S$-matrix element
\eqref{5.17}. The vectors \eqref{5.25} are spanned by the basis
vectors $|\bs{\p}j_3[\sm_R j]^-\>$ defined by \eqref{5.23}. 
This space of vectors \eqref{5.24} is labeled by the complex generalized
eigenvalue $\sm_R$ ($S$-matrix pole position) and the angular momentum
$j$ of the partial wave in which the resonance occurs. This means the
space of superpositions \eqref{5.24} is very similar to Wigner's
unitary representation spaces of the group of Poincar\'e
transformations for stable relativistic particles, the only difference
being that Wigner's
representation spaces are characterized by real square mass $m^2$ and
by spin $j$ $[m^2j]$, whereas the spaces of \eqref{5.24} are labeled by the complex number
$\sm_R$ and by $j$, the total angular momentum
of the scattering system of the $j$-th partial wave $a_j(\sm)$.

The association \eqref{5.24} between representation spaces $[\sm j]$ and
partial wave amplitude $a_j^{BW}(\sm)$
required a very 
specific form for the partial wave amplitude, namely the one given by the 
Cauchy kernel \eqref{5.22}.  
Similar associations of vectors to a partial wave amplitude
will not be possible if the partial wave amplitude has not the special
form of \eqref{5.22}~\footnote{For
instance, it would not be possible for the most popular form of a
relativistic Breit-Wigner with an energy dependent width given by the
on-the-mass-shell renormalization scheme \cite{tR98,passera}.}.
Even for the Breit-Wigner 
\eqref{5.22} we had to extend the values of $\mathsf s$ from the 
phenomenologically testable values $\sm_0\leq {\mathsf s}<\infty$ 
to the negative axis and introduce the idealization of an
``exact'' Breit-Wigner \eqref{5.22} for which $\mathsf s$ extends over 
$-\infty _{II}<{\mathsf s}<+\infty$.  
Only for the exact Breit-Wigner 
\eqref{5.22} could we use the Titchmarsh theorem
in \eqref{5.21}, \eqref{5.27} and associate to the 
amplitude $a_j^{BW}({\mathsf s})$ a vector which is defined by this exact 
Breit-Wigner amplitude.  And in order to apply the Titchmarsh theorem we had
to restrict the admissible wave functions $\overline{\psi}^-({\mathsf
s})=\<^-\psi|\sm^-\>$ and $\phi ^+({\mathsf s})=\<^+\sm|\phi^+\>$ 
to be Hardy class in the lower half plane \eqref{4.29}. 
That means we had to specify the in-state vector $\phi ^+$ and the 
out-observable vector $\psi ^-$ that can appear in the $S$-matrix 
element to be in the spaces 
$\Phi _-$ and $\Phi _+$, respectively.  
Only then could we 
define the Gamow kets $|[\sm_Rj]\eta^-\>$ in terms of the 
Dirac-Lippmann-Schwinger kets $|[\sm j]\eta^-\>$  
by, \eqref{5.25} or \eqref{5.27}, as functionals over the Hardy  space $\Phi _+$.
The Gamow vectors cannot  be defined as functionals
over the Schwartz space $\Phi$ like the usual Dirac kets. 
Thus the Hardy spaces $\Phi _-$ 
and $\Phi _+$, and therewith the new hypothesis \eqref{4.29},
\eqref{4.36}
had to be introduced (as in the non-relativistic 
theory \cite{aB97}) in order to be able to construct vectors
\eqref{5.23}, \eqref{5.25} 
with a Breit-Wigner energy distribution.

Similarly to the Gamow vectors \eqref{5.23} and \eqref{5.25}, 
we can define another kind of Gamow ket 
$|[\sm^*_Rj]\eta^+\>\in\Phi _-^{\times}$
in terms of the Dirac-Lippmann-Schwinger kets $|[\sm j]\eta^+\>$ 
for the $S$-matrix pole at ${\mathsf s}_R^*=(M_R+i\Gamma /2)^2$ in the upper 
half plane of the second sheet. We do not want to discuss this here.

From the relativistic Gamow vectors we can now calculate
consequences without any further mathematical assumption. This means
they are just consequences of the hypothesis \eqref{4.29} \eqref{4.36}
for the relativistic Gamow vectors defined 
as elements of $\Phi_+^\times$ from the $S$-matrix pole.

The relativistic Gamow vector $|\amd\>$
is a generalized eigenvector of $P^{\mu}$ with 
a complex eigenvalue.
To see this, we use the Titchmarsh theorem \eqref{5.27}
for the vector $\psi'^-\equiv P_{\mu}\psi^-\in\Phi_+$:
\begin{align}
\nonumber
\<P_\mu\psi^{-}\,|\,\amd\>
&=\frac{i}{2\pi}\int_{-\infty}^{\infty}d\sm\frac{
\<P_\mu\psi^{-}\,|\,\amr\>}
{\sm-\sm_{R}}\\
\nonumber
&=\frac{i}{2\pi}\int_{-\infty}^{\infty}d\sm
\frac{\sqrt{\sm} \p_\mu
\<\psi^{-}\,|\,\amr\>}
{\sm-\sm_{R}}\\
\label{5.28}
&=\sqrt{\sm_R}\p_\mu
\<\psi^{-}\,|\,\amd\>\, .
\end{align}
In \eqref{5.28},
we used \eqref{4.33a} to write 
$\<P_\mu\psi^-|\am\>=\sqrt{\sm}\p_\mu\<\psi^-|\am\>$ 
and~\eqref{4.32} to assert that $\sqrt{\sm}\p_\mu\<\psi^-|\am\>$
is again a Hardy  function from below, so that
Titchmarsh theorem (\eqref{h:1}, Appendix~\ref{h}) can be applied to this Hardy function
to obtain the last equality.
Similarly, for the square mass operator $M^2=P_\mu P^\mu$ we calculate 
\begin{equation}
\label{5.29}
\begin{split}
\<M^{2}\psi^{-}\,|\,\amd\>
&=\frac{i}{2\pi}\int_{-\infty}^{\infty}d\sm\frac{
\<M^{2}\psi^{-}\,|\,\amr\>}
{\sm-\sm_{R}}\\
&=\frac{i}{2\pi}\int_{-\infty}^{\infty}d\sm
\frac{\sm
\<\psi^{-}\,|\,\amr\>}
{\sm-\sm_{R}}\\
&=\sm_R
\<\psi^{-}\,|\,\amd\>\, .
\end{split}
\end{equation}
Equation \eqref{5.29}, valid for all $\psi^-\in\Phi_+$,
is the mathematical expression that $|\amd\>$ is a generalized eigenvector
of the square mass operator 
$M^{2}$ with the complex eigenvalue $\sm_R$. This 
is written equivalently as
\begin{equation}
\label{5.30}
M^{2^{\times}}|\amd\>=\sm_R|\amd\>\,.
\end{equation}

In the same way one can calculate the eigenvalue of the spin operator
\begin{equation}
\hat{W}=-\hat{w}_\mu\hat{w}^\mu,\ \text{with}\  
\hat{w}^\mu=\epsilon^{\mu\nu\rho\sigma}\hat{P}_\nu
J_{\rho\sigma}\nonumber
\end{equation}
where $\hat{P}_\mu=P_\mu M^{-1}$. One obtains, just in the same way as
for the basis vectors of Wigner's unitary representations 
\begin{equation}
\hat{W}|\bs{\p}j_3[\sm j]^-\>=j(j+1)|\bs{\p}j_3[\sm j]^-\>\nonumber
\end{equation}
This means that the numbers $[\sm_Rj]$ that label the spaces of Gamow
vectors are indeed the eigenvalues of square mass and spin, only that
this generalized eigenvalue $\sqrt{\sm}$ of the mass operator for
the relativistic Gamow vectors is now a complex number. The
representation spaces $[\sm_Rj]$ for resonances is an eigenspace (of
generalized eigenvectors) of the mass and spin operators
\begin{subequations}
\label{5.31}
\begin{equation}
(P^\mu P_\mu)^\times|[\sm_Rj]f^-\>=\sm_R|[\sm_Rj]f^-\>\label{5.31a}
\end{equation}
\begin{equation}
\hat{W}|[\sm_Rj]f^-\>=j(j+1)|[\sm_Rj]f^-\>\label{5.31b}
\end{equation}
\end{subequations}
in complete analogy to the Wigner representation spaces $[m^2j]$ for
stable particles. 

In the subsequent paper we shall derive physically
important properties of these representation spaces
characterized by $[\sm_Rj]$ and spanned by the Gamow kets \eqref{5.23}.
We shall show that, in contrast to Wigner's unitary representations $[m^2j]$ for
stable particles, the representations $[\sm,j]$ are 
{\it not} irreducible representation  of the
Poincar\'e {\em group} but  representation spaces of the subsemigroup 
into the forward light cone.
This will lead to causal propagation of Born probabilities.

In the subsequent paper, we shall also discuss the definition of 
resonance mass and width of a relativistic resonance.
That the complex number $\sqrt{\sm_R}$, defined by the $S$-matrix pole
position, characterizes a relativistic resonance does not yet tell us
how one should parameterize this complex number in terms of two real
numbers which could be conveniently called resonance mass and
width. One has an infinite number of possibilities 
to express the complex number $\sqrt{\sm_R}$ in terms of resonance 
mass and width of which some popular suggestions are \cite{tR98,passera,lC98}
\begin{equation}
\sqrt{\sm_R}=\left(M_R-i\Gamma_R/2\right)
=\sqrt{\bar{M_Z}^2-i\bar{M_Z}\bar{\Gamma_Z}}
=\sqrt{\frac{m_1^2-im_1\Gamma_1}{1+\frac{\Gamma_1}{m_1}}}\label{5.32}
\end{equation}
 The lineshape of the resonance will not be sufficient to
discriminate between them. The transformation property for the Gamow
vectors, considered as state vectors
of the resonance will allow us to choose precisely the values 
 $M_R$ and $\Gamma_R$ as the mass and width of a relativistic resonance.

\section*{Acknowledgement}
We gratefully acknowledge support from the Welch Foundation.
\section*{Appendices}
\appendix
\section{Overview of Rigged Hilbert Space Concepts}\label{r}
A Rigged Hilbert Space~\cite{gelfand} is the result of the completion
of a scalar product space with respect to three different topologies.
The completion of a vector space with respect to some topology
$\tau$ amounts to including in this space the limit points
of all $\tau$-Cauchy sequences. If one starts with a 
scalar-product  space $\Psi$ and completes it with respect
to the norm induced by the scalar product
$$
\|\phi\|=\sqrt{(\phi,\phi)}\,,
$$
a Hilbert space $\H$ is obtained. On the other hand, if 
$\Psi$ is completed with respect to a topology $\tau_\Phi$
defined by a countable number of norms with some qualifications, 
a countably normed space $\Phi$ is obtained. This countably normed
topology $\tau_\Phi$ is finer than the Hilbert space 
topology $\tau_\H$ so that 
there are more 
$\tau_\Phi$-neighborhoods
than $\tau_\H$-neighborhoods. Hence:
$$
\Psi\subset\Phi\subset\H\,.
$$
A third space of interest is the space of antilinear functionals
on $\Phi$, denoted by $\Phi^\times$. Since $\Phi\subset\H$, it follows
that $\H^\times\subset\Phi^\times$. But, from Hilbert space
theory, $\H=\H^\times$. 
Hence
\begin{equation}
\label{r1}
\Phi\subset\H\subset\Phi^\times\,.
\end{equation}
The triplet~\eqref{r1} is called a Rigged Hilbert Space when 
$\Phi$ is nuclear and dense in $\H$ (with respect to $\tau_\H$).
The fundamental physical axiom of the Rigged Hilbert Space formulation
of quantum physics is that the set of states of the physical system
do not inhabit the entire Hilbert space $\H$ but an appropriately
defined dense subspace $\Phi$ of $\H$. The countably normed topology
of $\Phi$ is constructed so as to yield the algebra of relevant physical
observables continuous as mappings on $\Phi$. It is this feature
of Rigged Hilbert Space theory that is made use of in Section~\ref{sec2}
in making the distinction between the set of
prepared states $\Phi_{-}$ and registered observables $\Phi_{+}$
by taking $\Phi_\pm$ as dense subspaces of the same 
Hilbert space $\H$ as in~\eqref{phi+} \eqref{phi-}. This distinction
is what allows semigroup time evolution
 to be incorporated into the quantum mechanical
theory.

The action of an element $F\in\Phi^\times$ on $\phi\in\Phi$,
$F(\phi)$, is denoted--in the Dirac bra-ket notation--by
$$F(\phi)=\<\phi|F\>\,.$$
Since $\H\subset\Phi^\times$, it follows that the Dirac bra-ket
$\<\phi|F\>$ is an extension of the Hilbert space scalar
product in the sense that
$$\<\phi|F\>=(\phi,F)\quad\text{ for }\quad F\in\H\,.$$
The topology on $\Phi^\times$, denoted by $\tau_{\Phi^\times}$,
is the weak$^*$-topology induced by $\Phi$ on $\Phi^\times$.
This means that convergence in $\Phi^\times$ is defined by
\begin{equation}
\label{r2}
F_i\xrightarrow{\tau_{\Phi^\times}}F
\Longleftrightarrow\<\phi|F_i\>\rightarrow\<\phi|F\>\,,\quad
\text{ for all }\phi\in\Phi\,.
\end{equation}

To every $\tau_\Phi$-continuous operator $A$ on
$\Phi$, there corresponds a $\tau_{\Phi^\times}$-continuous operator
$A^\times$ defined on $\Phi^\times$ by 
\begin{equation}
\label{r3}
\<\phi|A^\times F\>\equiv\<A\phi|F\>\,,\quad\text{ for all }
\phi\in\Phi\,,\quad F\in\Phi^\times\,.
\end{equation}
The operator $A^\times$ is called the conjugate 
operator of $A$. It is an extension of the 
Hilbert space adjoint operator $A^\dagger$, since for $F\in\H$ we have
\begin{equation}
\label{r4}
\<\phi|A^\times F\>=(A\phi,F)=(\phi,A^\dagger F)\quad\text{for }
F\in\H\,.
\end{equation}
Hence,
\begin{equation}
\label{r5}
A^\dagger|_{\Phi}\subset A^\dagger\subset A^\times\,.
\end{equation}
It should be stressed that the conjugate operator $A^\times$
can be defined as a $\tau_{\Phi^\times}$-continuous  
operator only when $A$ is a
continuous linear operator on $\Phi$.
In quantum mechanics, it is impossible (empirically) to restrict
oneself to continuous (and therefore bounded) operators $\bar{A}$
in $\H$. However, one can restrict oneself to algebras of observables
$\{A,B,\cdots\}$ described by continuous operators in $\Phi$, if
the topology of $\Phi$ is suitably chosen. Then, $A^\times$, $B^\times$,
$\cdots$ are defined and continuous in $\Phi^\times$.

A generalized eigenvector $|F\>$ of a $\tau_\Phi$-continuous
operator $A$ with a generalized eigenvalue $\omega\in{\mathbb C}$
is defined by the relation
\begin{equation}
\label{r6}
\<A\phi|F\>=\<\phi|A^\times F\>=\omega\<\phi|F\>\,,\quad
\text{ for all }\phi\in\Phi\,.
\end{equation}
Since the vector $\phi$ in~\eqref{r6} is arbitrary, \eqref{r6} can be
formally expressed as
\begin{equation}
\label{r7}
A^\times|F\>=\omega|F\>\,.
\end{equation}
In the Dirac notation the $^\times$ in~\eqref{r7} is suppressed
so that \eqref{r7} reads
\begin{equation}
\label{r8}
A|F\>=\omega|F\>\,.
\end{equation}
If $A$ is a self-adjoint operator, suppressing the $^\times$ 
as in~\eqref{r8} does not lead to confusion since 
$A=A^\dagger\subset\Phi^\times$.
However, if  $A$ is not self-adjoint, a clear distinction between
the operator and its conjugate should be made.
The concept of generalized eigenvectors~\eqref{r7} in  
Rigged Hilbert Space mathematics allows the description of ``eigenstates''
which do not exist in the Hilbert space. For instance, the Dirac scattering 
kets are generalized eigenvectors with eigenvalues belonging to the 
continuous spectrum, and they are not Hilbert space elements. The
Gamow vectors, 
which are used to describe decaying states, are also generalized
eigenvectors which are not in $\H$, 
but, unlike in the case of scattering states, 
their complex eigenvalues do not belong to the Hilbert
space spectrum of the Hamiltonian.
\section{Hardy Class Functions on a Half-plane}\label{h}
\begin{definition}[$\H_\pm^p$ $1\leq p<\infty$]\label{h:1}\cite{hardy,koosis}\cite{aB97} Appendix.
A complex function $f(x+iy)$ analytic in the open lower
half  complex plane $(y<0)$ is said to be a Hardy class function from below of order $p$, $\H_-^p$, if
$f(x+iy)$ is $L^p$-integrable as a function of $x$ for any $y<0$ and
\begin{subequations}
\begin{equation}
\label{h1}
\underset{y<0}{\rm sup}\int_{-\infty}^{\infty}
dx\,\,|f(x+iy)|^{p}<\infty\,.
\end{equation}
Similarly, a complex function $f(x+iy)$ analytic in the open 
upper half complex plane $(y>0)$ is said to be a Hardy
class function from above of order $p$, $\H_+^p$, if
$f(x+iy)$ is $L^p$-integrable as a function of $x$ for all $y>0$, and
\begin{equation}
\label{h2}
\underset{y>0}{\rm sup}\int_{-\infty}^{\infty}
dx\,\,|f(x+iy)|^{p}<\infty\,.
\end{equation}
\end{subequations}
\end{definition}
A property of $\H_\pm^p$ functions is that
their boundary values on the real axis exist almost everywhere and
define an $L^p$-integrable function, i.e.,
if $f\in\H_\pm^p$, then its boundary values 
$f(x)\in L^p({\mathbb{R}})$. Conversely, the values of any
$\H_\pm^p$ function on the upper/lower half-plane are 
determined from its boundary values on the real axis. This result
is provided by a theorem of Titchmarsh:
\begin{theorem}[Titchmarsh theorem]\label{a.1}
If $f\in\H_-^p$,  then 
$$
f(z)=\frac{-1}{2\pi i}\int_{-\infty}^{\infty}
\frac{f(t)}{t-z} dt\,,\quad\text{ for }{\rm Im}\,z<0\,,
$$
and
$$
\int_{-\infty}^{\infty}\frac{f(t)}{t-z}dt=0\,,\quad\text{ for }{\rm Im}\,z>0\,.
$$
Similarly, if $f\in\H_+^p$, then
$$
f(z)=\frac{1}{2\pi i}\int_{-\infty}^{\infty}
\frac{f(t)}{t-z} dt\,,\quad\text{ for }{\rm Im}\,z>0\,,
$$
and
$$
\int_{-\infty}^{\infty}\frac{f(t)}{t-z}dt=0\,,\quad\text{ for }{\rm Im}\,z<0\,.
$$
\end{theorem}
This one-to-one correspondence between the $\H_\pm^p$ functions and
their boundary values on ${\mathbb R}$ allows the identification
of $f(z)$ with $f(x)$ for $f\in\H_\pm^p$.

The following results are related to the decay properties
of the Hardy class functions. They are straightforward
generalizations of the corresponding results of~\cite{gadella}
and are needed for the construction of the relativistic Gamow vectors.
\begin{proposition}\label{h:2}
Let ${C}_{\infty}$ be the infinite semi-circle in the lower
half complex plane. If $f\in\H_-^p$, then
$$
\int_{{\cal C}_\infty}\left|\frac{f(z)}{z} dz\right|=0\,.
$$
\end{proposition}
\begin{proof}
Let ${\cal C}_r$ be the arc with radius $r$ shown in 
Figure~\ref{arc}. Then
$$
\left|\int_{{\cal C}_r}\frac{f(z)}{z} dz\right|\leq 
\int_{{\cal C}_r}|f(r e^{i\theta})|d\theta
=\int_{1/r}^{\pi-1/r}|f(-r e^{i\theta})|d\theta\,.
$$
Since $f\in\H_-^p$, then there exists $C$ such that
$$
|f(-re^{i\theta})|\leq\frac{C}{(r\sin \theta)^{1/p}}\,,
\quad\text{(cf.~\cite{koosis} page 149)}\,.
$$
Thus
\begin{equation}
\label{h3}
\int_{{\cal C}_r}\left|\frac{f(z)}{z} dz\right|\leq 
\frac{2C}{r^{1/p}}\int_{1/r}^{\pi/2}
\frac{1}{(\sin\theta)^{1/p}}d\theta\,.
\end{equation}
Using 
$$\sin\theta\geq \theta-\theta^{3}/6
\geq\theta(1-\pi^2/24)\,,\text{ for }
1/r\leq\theta\leq\pi/2\,,
$$
we obtain for~\eqref{h3}
\begin{eqnarray}
\lefteqn{\int_{{\cal C}_r}\left|\frac{f(z)}{z} dz\right|\leq
\frac{2C}{r^{1/p}(1-\pi^2/24)^{1/p}}
\int_{1/r}^{\pi/2}\frac{d\theta}{\theta^{1/p}}}\nonumber\\
& &\!\!\!\!=
\frac{2C}{(1-\pi^2/24)^{1/p}r^{1/p}}
\begin{cases}
\log\left(\frac{r\pi}{2}\right)
&\,\,\, p=1\\
\frac{1}{1-\frac{1}{p}}
\left[(\frac{\pi}{2})^{1-1/p}-(\frac{1}{r})^{1-1/p}\right]
&\,\,\, 1<p<\infty 
\end{cases}
\end{eqnarray}
Therefore, as $r\rightarrow \infty$, we obtain
$$
\int_{{\cal C}_\infty}
\left|\frac{f(z)}{z} dz\right|=0\,.
$$
\end{proof}
\begin{corollary}\label{h:3}
Let $f\in{\cal S}\cap\H_-^2$,
$g\in{\cal S}\cap\H_-^2$, then
$$
\int_{{\cal C}_\infty}|f(z)g(z) dz|=0\,.
$$
\end{corollary}
\begin{proof}
Since $f\in{\cal S}\cap\H_-^2$, then
$xf(x)\in{\cal S}\cap\H_-^2$~\cite{gadella}. A 
straightforward application of H\"older's inequality
shows that $xf(x)g(x)\in\H_-^1$. Then, from the above lemma
$$
\int_{{\cal C}_\infty}|f(z)g(z)dz|
=\int_{{\cal C}_\infty}\left|\frac{zf(z)g(z)}{z} dz\right|=0\,.
$$
\end{proof}

A remarkable property of Hardy class functions that is
used in~\cite{gadella} is that they are uniquely determined 
from their boundary values on a semi-axis on the real line.
This result is provided by a theorem of van Winter~\cite{winter}. 
Before stating the van Winter's theorem below, we define first
the Mellin transform
\begin{definition}[Mellin transform]
Let $f(x)$ be a function on ${\mathbb R}_{+}$. Its Mellin 
transform is a function defined almost everywhere on ${\mathbb R}$
as
$$
H(s)=\frac{1}{(2\pi)^{1/2}}\int_{0}^{\infty}f(x)x^{is-1/2}dx\,,
$$
provided that the integral exists for almost all $s\in{\mathbb R}$.
\end{definition}
\begin{theorem}[van Winter]\label{vanwinter}
A function $f(x)\in L^{2}({\mathbb R}^+)$ can be extended to
${\mathbb R}^{-}=(-\infty,0]$ to become a function in
$\H_{+}^{2}$ if and only if its Mellin transform satisfies
$$
\int_{-\infty}^{\infty}(1+e^{2\pi s})|H(s)|^{2}ds<\infty\,.
$$
This extension is unique. The values of $f(z)$ for 
$z=\rho e^{i\theta}$ for $0\leq \theta\leq \pi$, $\rho>0$
are given by
$$
f(\rho e^{i\theta})=\frac{1}{(2\pi)^{1/2}}\int_{-\infty}^{\infty}
H(s)(\rho e^{i\theta})^{-is-1/2}ds\,.
$$
In particular for negative values, the function is given by
$$
f(-x)=\frac{1}{(2\pi)^{1/2}}\int_{-\infty}^{\infty}
H(s)(x e^{i\pi})^{-is-1/2}ds\,.
$$
A similar result can be obtained for $\H_{-}^{2}$.
\end{theorem}

\newpage
\renewcommand{\thefigure}{1\alph{figure}}

\begin{comment}

\begin{figure}[p]
\centering
\subfigure{
\scalebox{.9}{\includegraphics{scatt_a.eps}}
}
\caption{Preparation of the controlled in-state.}
\label{scatt_fig_a}
\subfigure{
\scalebox{.9}{\includegraphics{scatt_b2.eps}}
}
\caption{The uncontrolled out-state: $\phi^{out}=S\phi^{in}$.}
\label{scatt_fig_b}
\end{figure}

\begin{figure}
\subfigure{
\scalebox{.9}{\includegraphics{scatt_c.eps}}
}
\caption{Registration of the detector defined out-observable.}
\label{scatt_fig_c}
\subfigure{
\scalebox{.9}{\includegraphics{scatt_d.eps}}
}
\caption{Combining the preparation of the state and the registration of the
observable in a scattering experiment.}
\label{scatt_fig_d}
\setcounter{figure}{0}
\renewcommand{\thefigure}{\arabic{figure}}
\caption{The subdivision of a quantum mechanical scattering experiment into a registration
part and a preparation part.}
\label{scatt_fig}
\end{figure}

\end{comment}

\begin{figure}[p]
\centering
\scalebox{.9}{\includegraphics{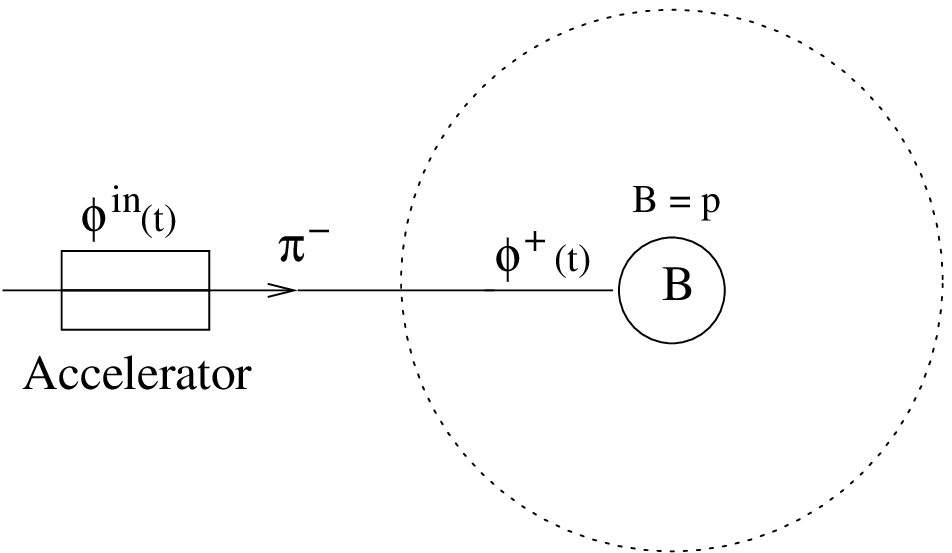}}
\caption{Preparation of the controlled in-state.}
\label{scatt_fig_a}
\end{figure}

\begin{figure}
\centering
\scalebox{.9}{\includegraphics{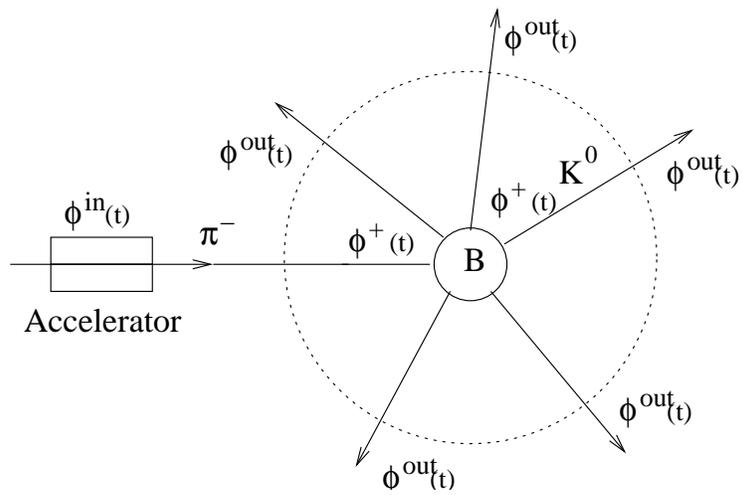}}
\caption{The uncontrolled out-state: $\phi^{out}=S\phi^{in}$.}
\label{scatt_fig_b}
\end{figure}

\begin{figure}
\centering
\scalebox{.9}{\includegraphics{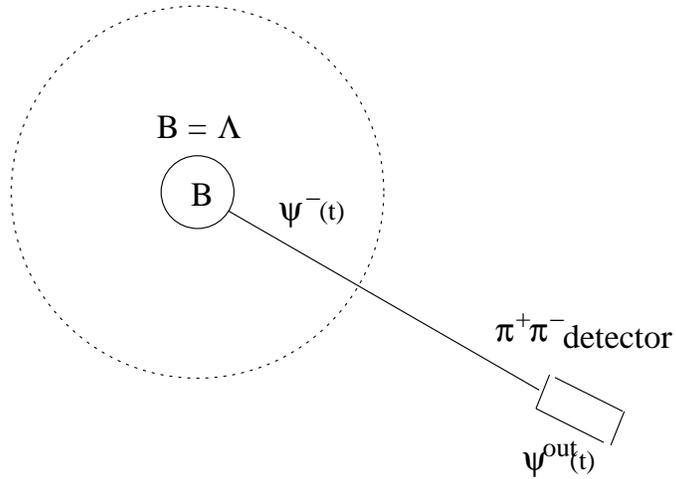}}
\caption{Registration of the detector defined out-observable.}
\label{scatt_fig_c}
\end{figure}

\begin{figure}
\centering
\scalebox{.9}{\includegraphics{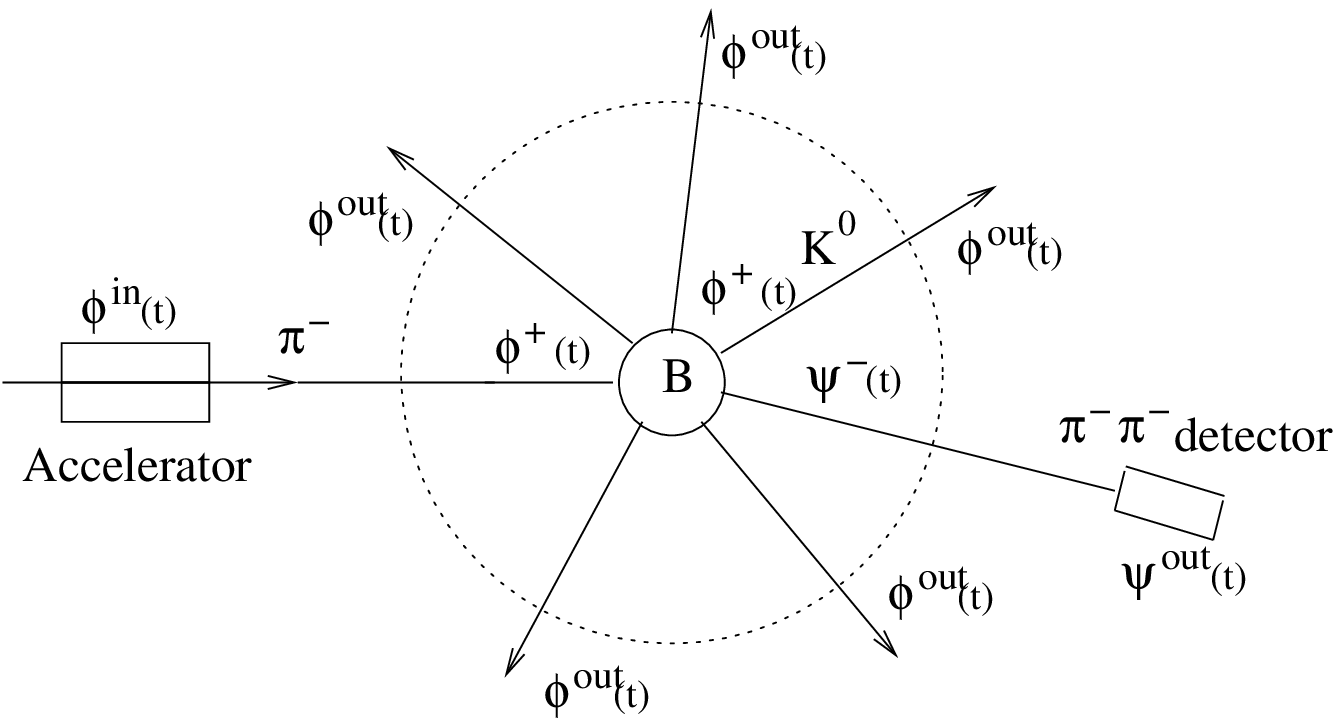}}
\caption{Combining the preparation of the state and the registration of the
observable in a scattering experiment.}
\label{scatt_fig_d}
\setcounter{figure}{0}
\renewcommand{\thefigure}{\arabic{figure}}
\caption{The subdivision of a quantum mechanical scattering experiment into a registration
part and a preparation part.}
\label{scatt_fig}
\end{figure}

\renewcommand{\thefigure}{\arabic{figure}}

\begin{figure}[p]
%\scalebox{.9}{\includegraphics{contour.eps}}
\begin{picture}(100,200)
\put(0,0){\includegraphics{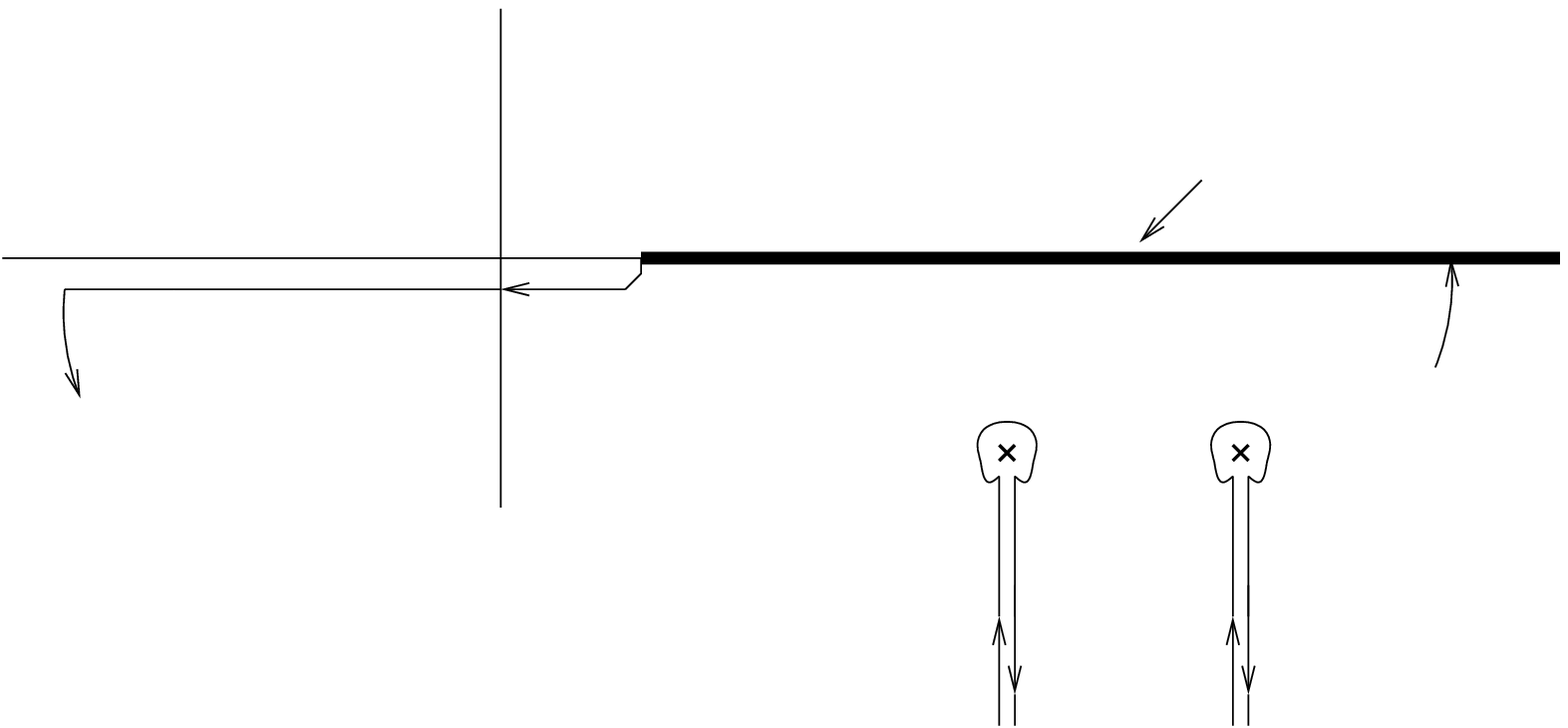}}
\put(0,100){$C_\infty$}
\put(180,145){$\sm_0$}
\put(337,165){Cut}
\put(400,100){$C_\infty$}
\put(280,95){$C_1$}
\put(303,80){$\sm_{R_1}$}
\put(350,95){$C_2$}
\put(373,80){$\sm_{R_2}$}
\put(60,110){$C_-$}
\put(150,205){$\sm$ First sheet}
\put(150,30){$\sm$ Second sheet}
\end{picture}
\caption{Contour of integration of~\eqref{5.11} after contour deformation. The figure shows
the second sheet of the $S$-matrix $S(\sm)$ with
the cut from $\sm_0$ to $\infty$, and two resonance poles at $\sm_{R_1}$ and
$\sm_{R_2}$ in the
second sheet. The upper half is the first sheet of $S(\sm)$ which one reaches
when one goes across the real axis between $\sm_0$ and $\infty$.
The original contour of integration of~\eqref{5.10} was along the cut.
This contour is then deformed into the contour shown: $C_-$, $C_\infty$, $C_1$, $C_2$, $C_\infty$
which leads to the integrals in \eqref{5.11}.}
\label{contour_fig}
\end{figure}

\begin{figure}[p]
\centerline{\scalebox{.9}{\includegraphics{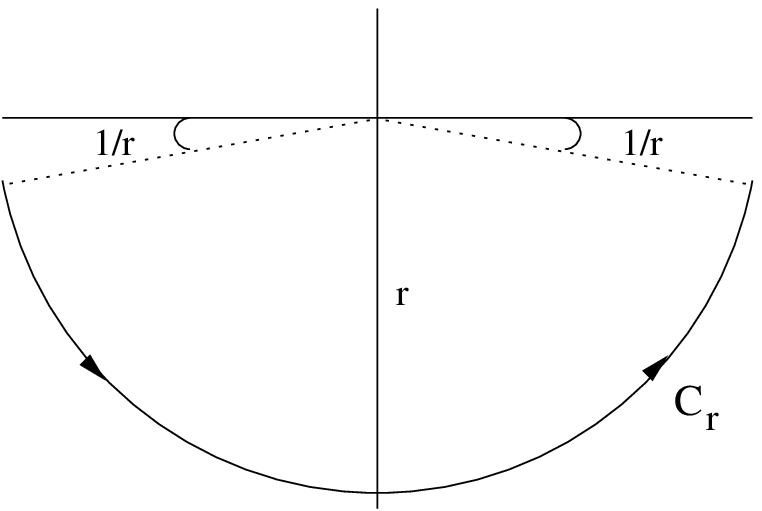}}}
\caption{(Proof of Proposition \ref{h:2} of Appendix \ref{h}). Arc ${\cal C}_{r}$.}\label{arc}
\end{figure}
\end{document}